\pdfoutput=1 
\documentclass{JINST}
\usepackage{multirow}

\title{Pixel area variations in sensors:  \\ a novel framework for predicting pixel fidelity and distortion in flat field response}

\author{Andrew Rasmussen\\
SLAC National Accelerator Laboratory,\\
  2575 Sand Hill Rd., Menlo Park, CA 94024\\
E-mail: \email{arasmus@slac.stanford.edu}}

\abstract{We describe the drift field in thick depleted silicon sensors as a superposition of a one-dimensional {\it backdrop} field and various three-dimensional {\it perturbative} contributions that are physically motivated. We compute trajectories for the conversions along the field lines toward the channel and into volumes where conversions are confined by the perturbative fields. We validate this approach by comparing predictions against measured response distributions seen in five types of fixed pattern distortion features. We derive a quantitative connection between "tree ring" flat field distortions to astrometric and shape transfer errors with connections to measurable wavelength dependence -- as ancillary pixel data that may be used in pipeline analysis for catalog population. Such corrections may be tested on DECam data, where correlations between tree ring flat field distortions and astrometric errors -- together with their band dependence -- are already under study. Dynamic effects, including the brighter-fatter phenomenon for point sources and the flux dependence of flat field fixed pattern features are approached using perturbations similar in form to those giving rise to the fixed pattern features. These in turn provide drift coefficient predictions that can be validated in a straightforward manner. Once the three parameters of the model are constrained using available data, the model is readily used to provide predictions for arbitrary photo-distributions with internally consistent wavelength dependence provided for free.}

\keywords{CCDs; pixel mapping; transverse drift; anisotropic diffusion; instrument signature removal}

\begin{document}

\section{Scope}\label{sec:a}

In this contribution we discuss modeling efforts that we formulated organically in response to a number of suspected {\it pixel size variation} related phenomena seen in LSST sensor prototype characterization data. By {\it modeling} we mean, ultimately, geometric simulation -- in the sense that electromagnetic effects in media, such as transmission, refraction and conversion of incoming light -- are handled upstream in a ray trace like fashion~\cite{Groom:1999,Rajkanan:1979} -- but that the apparent lateral diffusion via Brownian motion of the photo-conversions, occurs over the drift trajectory that connects those conversion positions to the potential wells of the buried channel where conversions are accumulated. 

Historically, we have dealt with the one-dimensional Poisson solution within depleted high resistivity silicon~\cite{Prigozhin:1998}, as a fundamental building block to any detailed photo-conversion registration models into the pixel grid array of potential wells. The following list of phenomena most easily seen in a sensor's response to flat field illumination, compel us to generalize from the one-dimensional field solutions to one where distributions of bound charges and/or equipotential surfaces perturb the "backdrop" 1-D solution:

\begin{enumerate}\itemsep0em
\item Tree rings;
\item Midline charge redistribution;
\item Edge rolloff;
\item Anti-correlations in collecting areas of adjacent pixels;
\item Effects associated with the onset of "tearing", and
\item Effects associated with pixel-to-pixel correlations in the number of conversions collected.
\end{enumerate}
In the sections that follow, we give plausible physical descriptions for the preceding effects together with predictive models that can be used to reproduce observable effects in the flat field response. We believe this is an important exercise to perform because the underlying model that reproduces flat field response, also predicts internally consistent sets of astrometric errors and shape transfer. 

Several authors~\cite{Smith:2008,Stubbs:thisvolume} argue compellingly that dividing bias- and dark current-corrected images through by flat field response is an erroneous step that proliferates in traditional CCD data reduction pipelines. We suggest here that the detail contained in flat field response distortions can be used instead to derive {\it ancillary pixel data} that can partially compensate for the astrometric and shape transfer errors that share the common origin described above - {\it the underlying mapping function}. Ancillary pixel data would include the following, each modestly dependent on wavelength:
\begin{enumerate}\itemsep0em
\item pixel geometric area error, or flat field response distortion;
\item pixel centroid error, or astrometric error (two values), and
\item pixel second moment error, or shape transfer (three values).
\end{enumerate}
Whether and how these corrections are used in any science analysis or catalog population, are subjects of a different study. For now, trivial (or null) values for each of these six ancillary (per) pixel data should converge with current, standard CCD image analysis that do not use such ancillary data.

\section{Distortion effects seen in flat field illumination}\label{sec:b}

\begin{figure}[tbp] 
\centering
\begin{tabular}{rr}
\includegraphics[height=.3\textwidth]{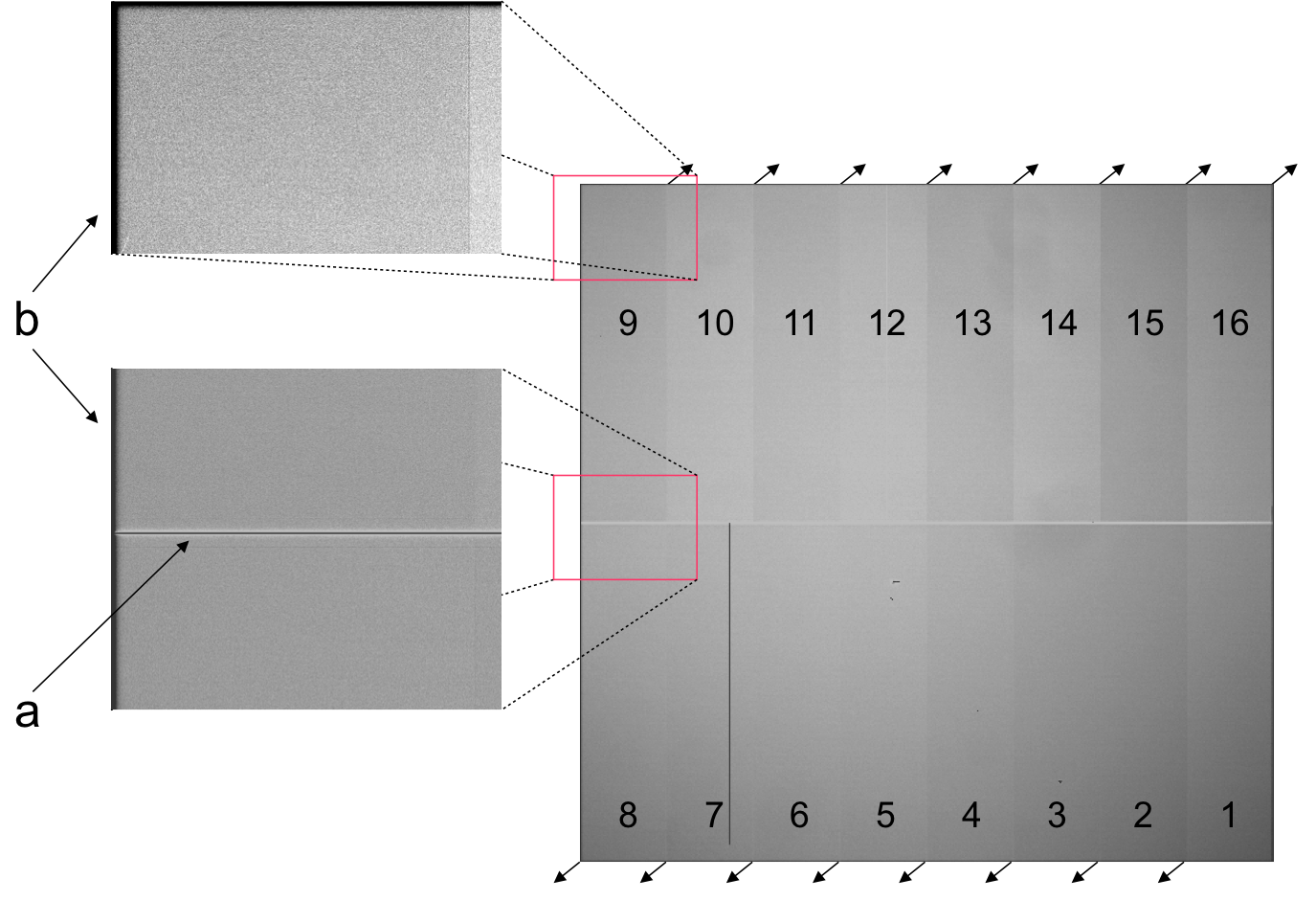}&
\includegraphics[height=.3\textwidth]{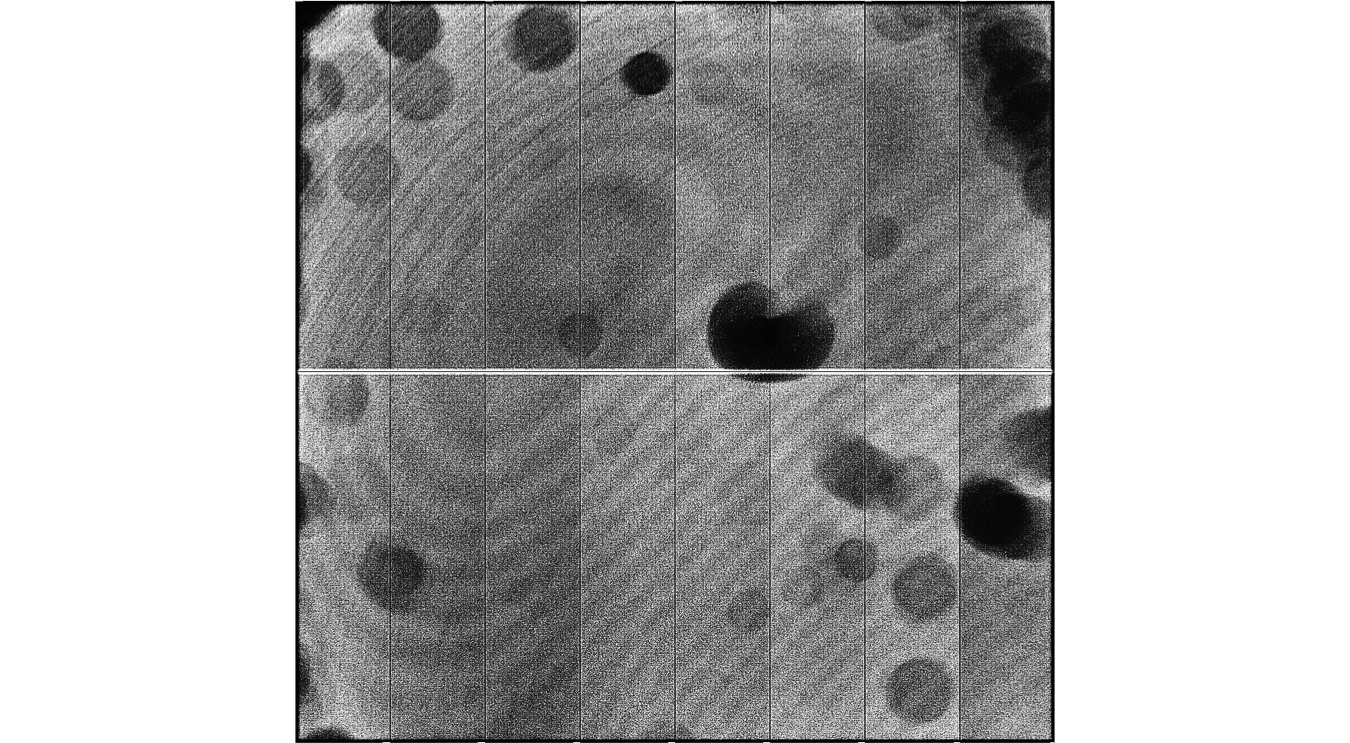}\\
\end{tabular}
\caption{An overview of the fixed pattern flat field distortions discussed here. Left: a reassembled image displayed with a stretch (light grey:dark grey) corresponding to roughly $\Delta \ln C \sim 0.3$. The midline charge redistribution feature is called out with the label "a", while the edge rolloff is labeled "b". Right: a similar image to that on the left, but this time the stretch corresponds to roughly $\Delta \ln C \sim 0.005$. The circular, dark spots are not of interest at this time - these may simply be due to projections of dust particles in the optical system, or real quantum efficiency variations due to AR coating layer thickness variations. Fringing contours are also not of interest, these are expected to be damped out strongly by the thick sensors that intercept LSST's F/1.2 beam. The fine striations with approximately 30 pixel period, roughly parallel to the fringing, are the flat field distortions referred to as tree rings.}
\label{fig:flatfield}
\end{figure}

Distortions seen in the flat field response of a sensor can be broadly categorized into two classes: {\it fixed pattern} and {\it dynamic}. Later we will show that this breakdown does not strictly hold true, but for now we use these labels. The first four in the list above we consider {\it fixed pattern}; the sixth one is {\it dynamic}. The fifth is a special case, because it is {\it too} dynamic and poorly understood: sensor operation prescriptions will be chosen that minimize or eliminate completely the phenomenon we refer to as {\it tearing}. For the purposes of this study though, we find some of the patterns seen under this category to be fundamental to the overall understanding of how drift field lines can be distorted by apparent changes in the bound charge configuration at the front side. Figure~\ref{fig:flatfield} gives an incomplete overview of the effects we discuss here.

\section{Tools}\label{sec:c}

Prior to describing each of the five effects listed above, it is important to write down equations that describe the drift field {\it backdrop} - which is equal to the unperturbed drift field and is a simple one-dimensional solution of the Poisson equation. The backdrop electric field strength $E^{BD}(z)=\vec{E}^{BD}\cdot\hat{z}$ can be expressed as the superposition of contributions by bound charges within the depleted Si, together with the field due to the external imposed "over depletion bias": $E^{BD}(z) = E^{bound}(z)+E^{ODB}$ where $E^{ODB} = V_{BSS}/t_{Si}$.\footnote{We assume here that $V_{BSS}=0$ when the device is just depleted. We expect for there to be an offset that relates $V_{BSS}$ to the backside bias $BSS$ for a given choice of parallel clock potentials - {\it cf.} Fig.~\ref{fig:treerings}.} The built-in field contribution can be written:\footnote{With p-type Si, $N_a > 0$ counts as $N(z)< 0$.}
\begin{equation}
E^{bound}(z) = {1 \over \epsilon_0\epsilon_{Si}} \int^{t_{Si}}_{z} dz\,N(z).
\end{equation}
With the exception of any bound charges near the back side surfaces or the higher doped regions near the front (n-channel or $p+$ implants), the total number of bound charges within the Si bulk of a $10\times 10\mu$m pixel should be on the order of $10^{4}$ (if $N_a \sim 10^{12}\,\mathrm{cm}^{-3}$).
Neglecting possible off-diagonal terms in the mobility tensor, drift times and lateral diffusion scales are calculated for the conversions, in this case electrons:\footnote{Note that since the mobility $\mu(E,T)$ is not constant, velocity saturation effects end up governing the diffusion scale, and in turn, it cannot be computed accurately using the minimum electric field strength along its trajectory. For the purpose of this study, mobility parameterizations of Jacoboni~\cite{Canali:1975PhRvB,Canali:1975ApPhL,Jacoboni:1977} were used.}
\begin{eqnarray}
t_{coll}(z) &=& \int^{z}_{z_{ch}} {dz \over \mu_e\left(E(z),T\right)\, \left| E(z) \right| }\\
\sigma^2(z) &=& {2kT \over q}\,\mu_e(E=0,T)\,t_{coll}(z).
\end{eqnarray}
Evaluating these expressions for our device characteristics and for typical operating parameters give typical collection times $t_{coll}\approx 1\mathrm{ns}$ and diffusion scales $\sigma\approx 4.2\mu\mathrm{m}$ for surface conversions. 

Inclusion of any three dimensional perturbations to the electric field, $\delta \vec{E}(\vec{x})$, such that the total drift field can be expressed as $\vec{E}^{tot}(\vec{x})=\vec{E}^{BD}(z)+\delta \vec{E}(\vec{x})$, results in a mapping, or a lateral shift of the conversions drifted toward the channel:
\begin{eqnarray}
\label{eqn:drift} \delta \vec{x}_\perp(\vec{x}_0) \cdot \hat{e}_{1,2} &=& \int^{\vec{x}_0}_{\vec{x}\cdot\hat{k}=z_{ch}} d\vec{l} \cdot \hat{e}_{1,2}\\
d \vec{l} &=& {\vec{E}(\vec{x}) \over \left| \vec{E}(\vec{x}) \right|}\, ds \\
t_{coll}(\vec{x_0}) &=& \int^{\vec{x}_0}_{\vec{x}\cdot\hat{k}=z_{ch}} {dl \over \mu_e\left(E(z),T\right)\, \left| E(z) \right| }.
\end{eqnarray}
In the above expression for lateral shift $\delta\vec{x}_\perp(\vec{x}_0)$, it is important to realize that the dimensions of the channel's potential well (along the serial transfer axis) is many times smaller than the pixel dimension - in other words, any trajectory that connects points within the photosensitive bulk ($\vec{x}_0$) to the channel ($\vec{x}\cdot \hat{k}=z_{ch}$) via the drift lines, is normally expected to produce a uniform range of lateral shifts such that $\left| \vec{x}_\perp(\vec{x}_0) \cdot \hat{i} \right| < 0.5$ pixels. What is more relevant is how positions near the {\it zero field saddle point locus}, or the pixel delimiter, are mapped via the drift lines, to positions $\vec{x}_0$ within the photosensitive bulk: $\delta\vec{x}_\perp(\vec{x}_0|\vec{x}_{sp})$. These define the effective shape of each pixel at the photo conversion depth. We find that with the tools in hand, forward propagation is the most reliable method for determining the position of the zero field saddle point locus, because it also responds sensitively to finite field perturbations $\delta\vec{E}(\vec{x})$.\footnote{The zero field, saddle point locus within the depleted Si are positions $\vec{x}_s$ with an electrostatic potential $\phi$ satisfying: $\vec{\nabla}\phi = 0$ {\it and} $\partial^2 \phi\partial z^2 < 0$ {\it and} ($\partial^2 \phi/\partial x^2 > 0$ or $\partial^2 \phi/\partial y^2 > 0$).}


\section{Quantitative comparisons of modeling results to data}\label{sec:d}

\subsection{Fixed pattern effects: tree rings}\label{sec:e}
Coherent, percent-level variation in flat field response is unlikely to be due to variations in quantum efficiency: wavelength dependence is seen in photon response non-uniformity (PRNU) curves. Unlike fringing, the phasing of the detailed response does not depend on wavelength, but the amplitude does (amplitudes decrease toward the near-IR). Tree ring amplitudes decrease in response to increased potential drop between backside bias and channel. Detailed flat field response is a manifestation of a mapping function that connects the pixel grid (at the channel) to corresponding coordinates at the backside window. The mapping function also predicts shape distortions of individual pixels (e.g., pixel elongation). Perpendicularity between astrometric errors and flat field response contours seen in DECam sky data confirm this straw man model for the underlying mechanism that make {\it tree rings} expressed in flat fields.

A variation in Si bulk resistivity is a fossil remnant of the process that was used to grow and purify the crystal ingot. Because of the cylindrical geometry this process was performed under, wafers made from the completed Si crystal will exhibit slices of that three dimensional concentric cylindrical arrangement. The variation in resistivity comes into play when the photosensitive volume is depleted of carriers - and regions of lower resistivity that correspond to a higher concentration of impurities - will possess a higher volumetric density of bound charge due to a higher concentration of acceptors ($N_a$) for example. Thus, solutions to Poisson's equation will inevitably be perturbed by such non-uniformities.

\begin{figure}[tbp] 
\centering
\begin{tabular}{rr}
\includegraphics[width=.35\textwidth,angle=-90]{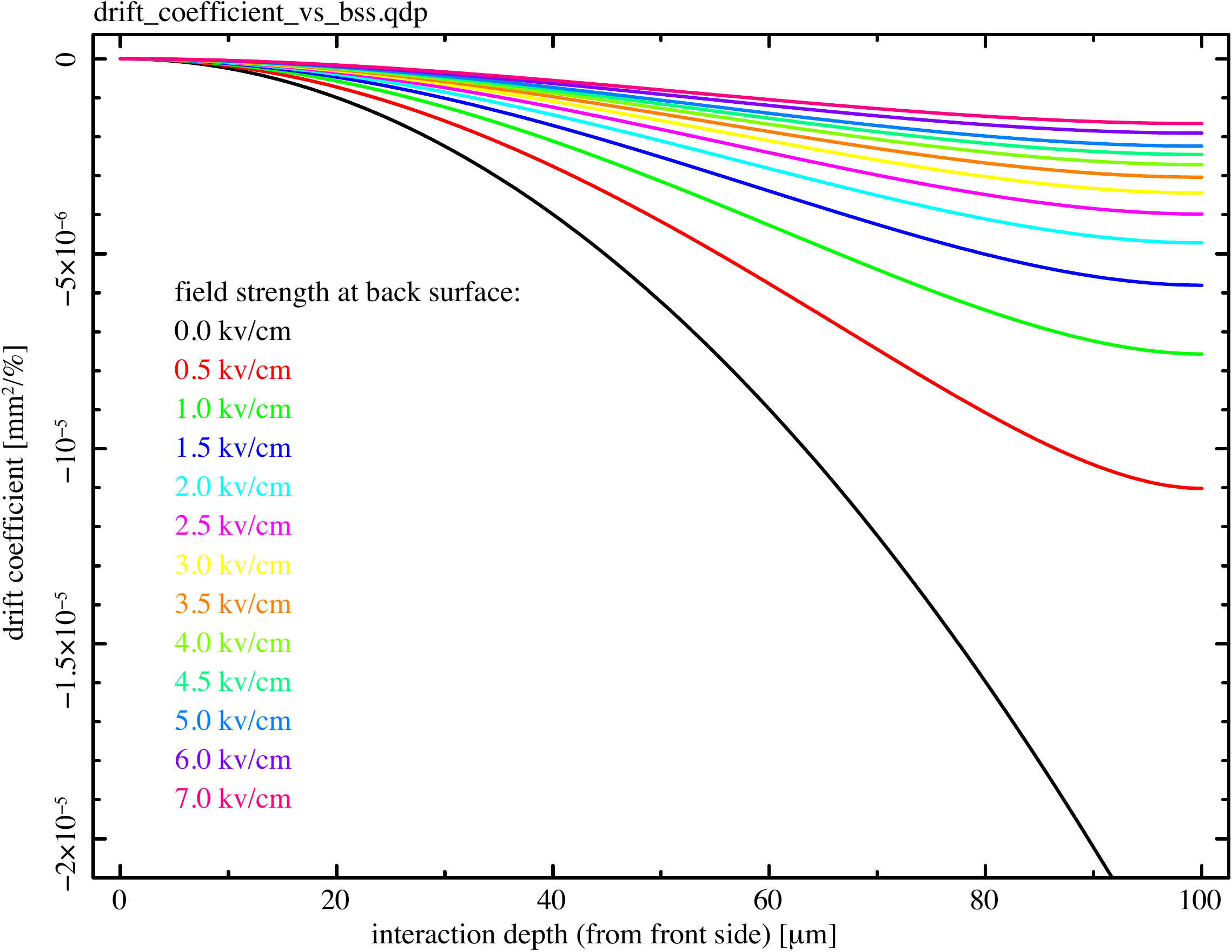}&
\includegraphics[width=.35\textwidth,angle=-90]{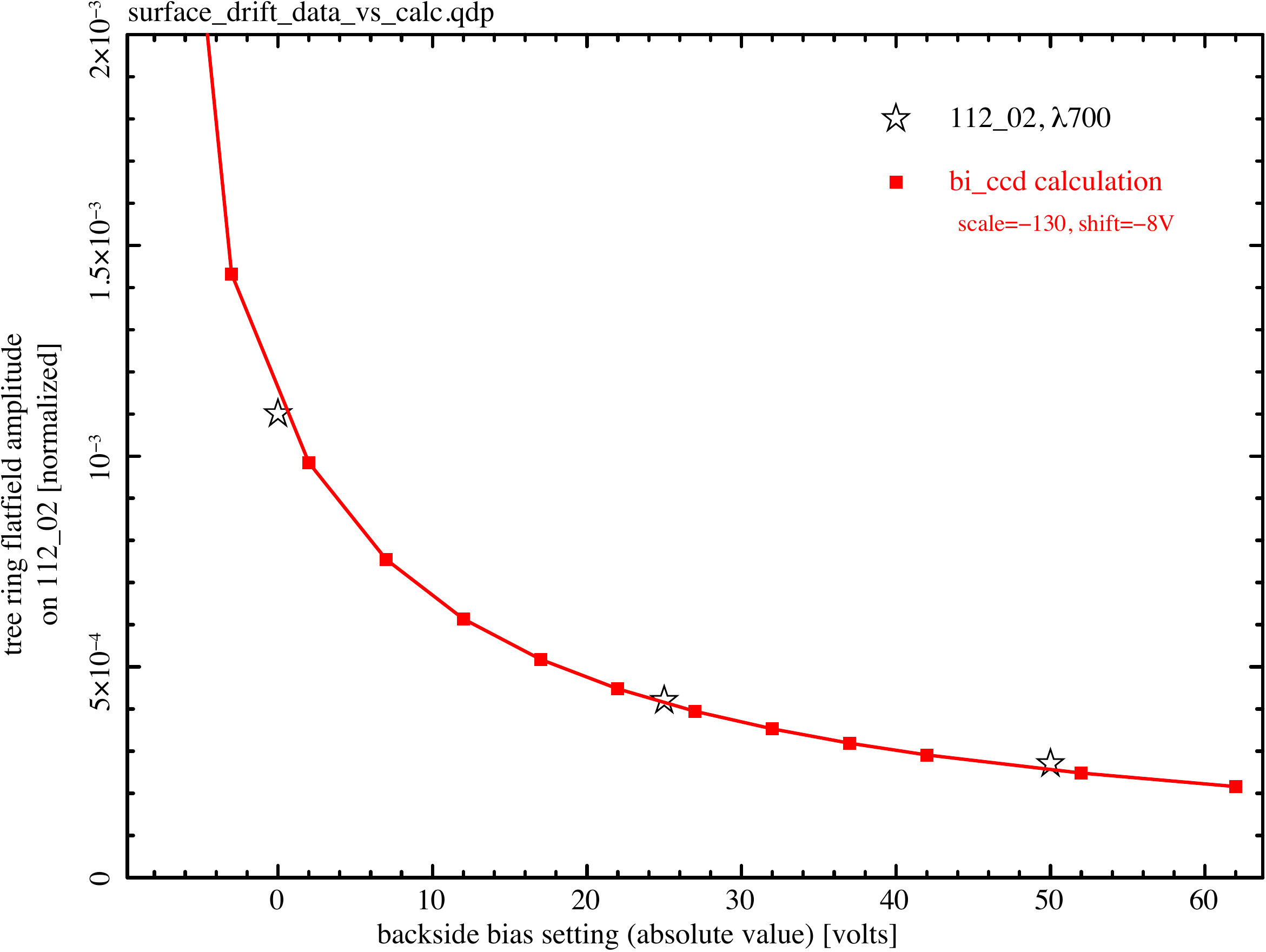}\\
\includegraphics[width=.35\textwidth,angle=-90]{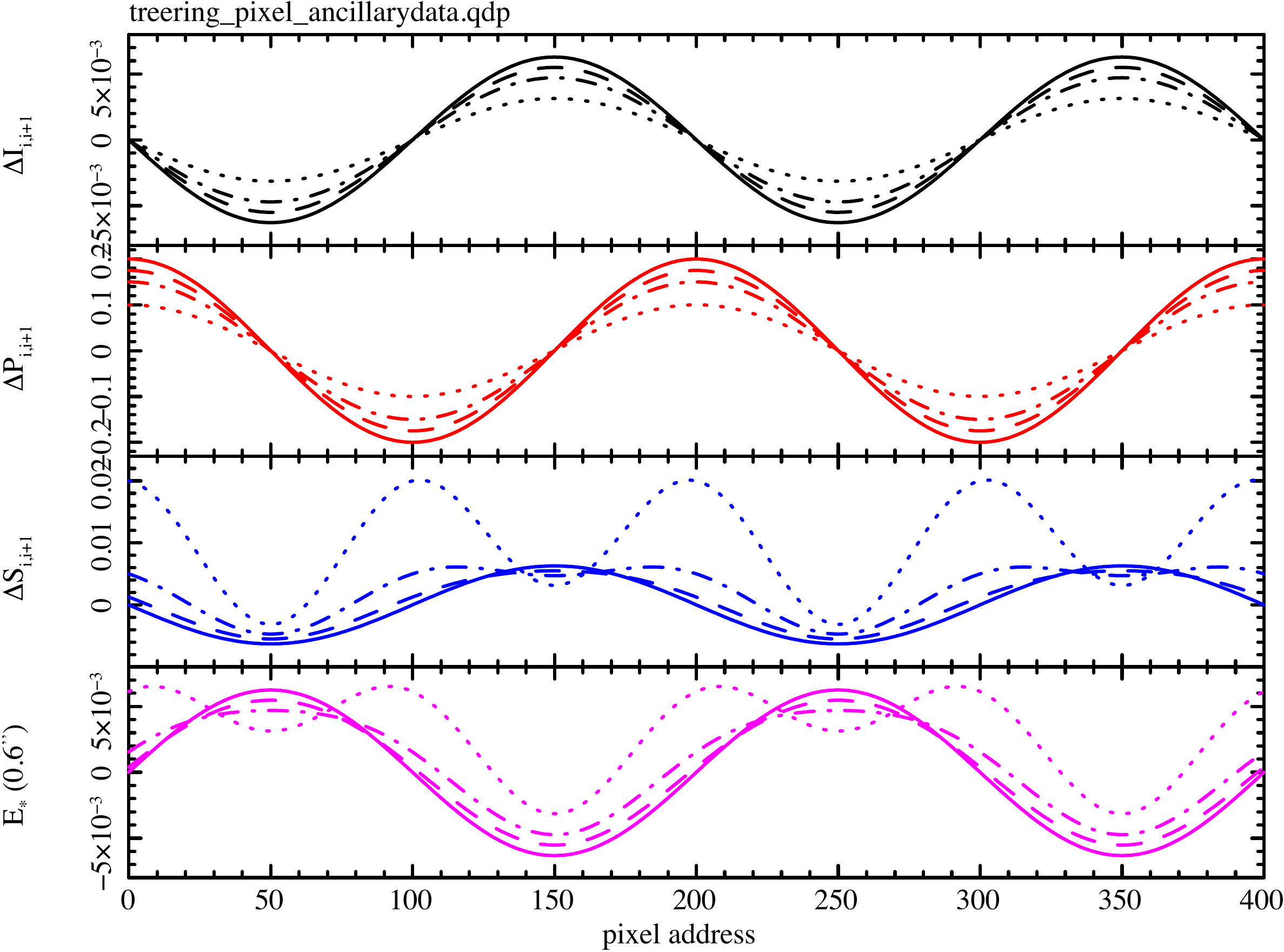}&
\includegraphics[width=.35\textwidth,angle=-90]{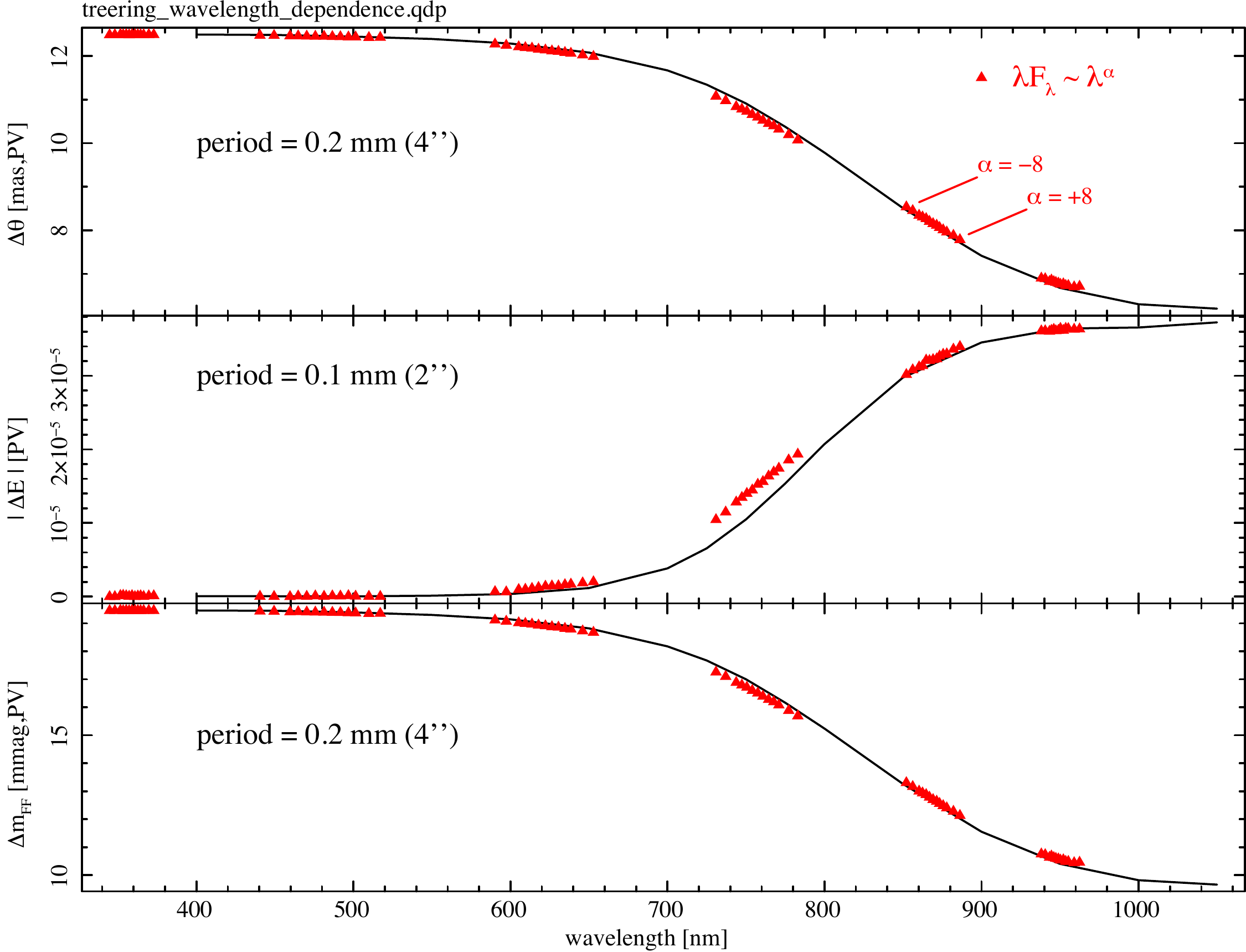}\\
\end{tabular}
\caption{Predictions for tree ring effects and a comparison to available characterization data. Upper left: drift coefficients $d\phi/dN_a$ for evaluated for different field strength values within the backside surface. Upper right: an alignment of the drift coefficient predictions for surface conversions, to three data points (stars) that were extracted from prototype device flat field distortion measurements. The scaling implies an electric field strength offset of 0.8keV/cm that corresponds to BSS=0V, and a impurity density amplitude of $\Delta N_a \sim 130 \mathrm{mm}^{-2}\,(p/2\pi)^2\times 10^{10}\,\mathrm{cm}^{-3} \approx 4\mathrm{E}9\,\mathrm{cm}^{-3}$. Lower left: analytic predictions for ancillary pixel data that would result from a sinusoidal flat field distortion pattern with PV amplitude of $\sim$1\% and a 200 pixel period. From the top tier toward the bottom, the functions plotted are $\Delta I_{i,i+1}$, $\Delta P_{i,i+1}$, $\Delta S_{i,i+1}$ and $E_\ast (0.6^{\prime\prime})$. The solid, dashed, dot-dashed and dotted line styles correspond to tree ring suppression factors ($\alpha$) of 0.0, 0.125, 0.25 and 0.5, respectively. Although it was not defined in the text, $E_\ast$ is the apparent, pixel response ellipticity ($\hat{e}_1$ component) of a perfectly isotropic source with 0.6 arc seconds FWHM. Lower right: a ray trace characterization of sample drift coefficients given in the upper left plot, showing the predicted wavelength dependence for the point response. Input assumptions include a "calibration point" of 14 millimag PV PRNU at 830nm at an observed period of 20 pixels. Relative astrometric error, shape transfer and flat field response are given. Pixelization effects have been neglected here so the shape transfer term excludes the term $\Delta S^0_{i,i+1}$.}
\label{fig:treerings}
\end{figure}

These perturbations can give rise to a non-trivial mapping function - and so variations in pixel sizes according to conversion depth. Using the definitions outlined above, specific solutions for the backdrop field $E^{BD}(z)$ are driven by the builtin field $E^{bound}(z)$ and the externally applied, over depletion bias field $E^{ODB}$. When there exists a gradient in the Si resistivity, neighboring regions can have different functions for $N(z)$, which in turn produce different builtin fields $E^{bound}(z)$. Because the backside window and the polysilicon gates form equipotential surfaces, the total potential drop $\Delta V$ should remain constant across regions with different $N(z)$. An approximation to the lateral field $\vec{E}_\perp(z) = \vec{E}(z)  - (\vec{E}(z)\cdot\hat{k}) \hat{k} $ can be constructed by aligning and subtracting adjacent solutions for $\phi(z) = \int^{t_{Si}}_{z} dz\,E^{BD}(z)$: $\vec{E}_\perp(z) \approx \left[\phi(z|N_z)-\phi(z|N^\prime_z)\right]/\Delta z = \left[ d\phi / dN_z \right] \, dN_z/dz$. Integration along the resulting drift field ({\it cf.} Eq. \ref{eqn:drift}) produces a drift coefficient curve which, when scaled by the local impurity density gradient, provides the depth dependence of the lateral drift. Figure~\ref{fig:treerings} (upper left) provides results of the drift coefficient calculation (essentially $d\phi/dN_z$). 

To connect lateral drift predictions to resulting distortions in the flat field response, consider a one-dimensional mapping function that is valid for surface conversions only, $\delta x(x|z=t_{Si})$. If we denote pixel boundaries along the x direction with subscripts $i$ and those along the perpendicular direction with subscripts $j$, mapping function can be used to generate the local flat field response (or relative pixel size) $\Delta I$, the astrometric shift of the pixel $\Delta P$, and the shape transfer or pixel ellipticity $\Delta S$ with:
\begin{eqnarray}
\Delta I^0_{i,i+1} &=& {\left[x_{i+1}+\delta x_{i+1})\right] - \left[x_{i}+\delta x_{i})\right] \over x_{i+1} - x_{i} } -1 =  {\delta x_{i+1} -\delta x_{i}\over x_{i+1}-x_{i}}  \\
\Delta P^0_{i,i+1} &=& {\delta x_{i+1} + \delta x_{i} \over 2} \\
\Delta S^0_{i,i+1} &\equiv& {I_{xx} - I_{yy} \over I_{xx}+I_{yy}} = {\left[(x_{i+1} + \delta x_{i+1}) - (x_{i} + \delta x_{i}) \right]^2 -\left[ y_{j+1} - y_{j} \right]^2 \over \left[(x_{i+1} + \delta x_{i+1}) - (x_{i} + \delta x_{i}) \right]^2 +\left[ y_{j+1} - y_{j} \right]^2} \approx \Delta I^0_{i,i+1}.
\end{eqnarray}
When a range of conversion depths is considered, the pixel boundary shift $\delta x_{i}$ will presumably be suppressed, on average, by a wavelength dependent factor $\alpha(\lambda)$. If we assume that the drift coefficient (in this case $d\phi/dN_z$) varies monotonically in $z$, an additional broadening of the pixel response will result from any non-zero suppression factor $\alpha$. These considerations, together with a uniform substitution of $\delta x_i$ with $(1-\alpha)\delta x_i$ and $I_{xx} \rightarrow {1\over 12} \left[(x_{i+1} + \delta x_{i+1}) - (x_{i} + \delta x_{i}) \right]^2 + {1\over 3} (\alpha\Delta P^0_{i,i+1})^2$, modify the above equations to form a new set of relations:
\begin{eqnarray}
\Delta I_{i,i+1} &=&  \left[ 1-\alpha(\lambda) \right] \Delta I^0_{i,i+1}  \\
\Delta P_{i,i+1} &=& \left[ 1-\alpha(\lambda) \right] \Delta P^0_{i,i+1} \\
\Delta S_{i,i+1} &=& \left[ 1-\alpha(\lambda)\right] \Delta I^0_{i,i+1} + 2 \left( {\alpha(\lambda)\Delta P^0_{i,i+1} \over x_{i+1} - x_{i}} \right)^2 = \Delta I_{i,i+1} + 2 \left( {{\alpha \over 1-\alpha }\Delta P_{i,i+1} \over x_{i+1} - x_{i}} \right)^2.
\end{eqnarray}
This analysis can readily be extended into the Fourier domain by considering a continuous mapping function that is valid at the pixel boundaries. Starting out by defining $\delta x_i \equiv \int dk\,a_k\,e^{ikx_i}$, $\tilde{x}\equiv (x_{i+1}+x_i)/2$ and $\Delta\equiv (x_{i+1}-x_{i})$, the following relations are obtained:
\begin{eqnarray}
\Delta I_{i,i+1}  &=& {(1-\alpha)\over \Delta} \int dk\,a_k\,2i\,e^{ik\tilde{x}}\,\sin(k\Delta/2)\\
\Delta P_{i,i+1} &=& (1-\alpha) \int dk\,a_k\,e^{ik\tilde{x}}\,\cos(k\Delta/2)\\
\Delta S_{i,i+1} &=& {1-\alpha \over\Delta} \int dk\,a_k\,2i\,e^{ik\tilde{x}}\sin(k\Delta/2) + 2 \left({\alpha\over\Delta}\int dk\,a_k\,e^{ik\tilde{x}}\,\cos(k\Delta/2)\right)^2
\end{eqnarray}

The apparent complexity in the expression for $\Delta S$ above should be noted: for finite tree ring suppression $\alpha$, the expression contains terms both from flat field distortion $\Delta I$ and its integral $\Delta P$. The latter term is squared, which means that it in any periodic variation in impurities, shape transfer response should be seen at both the intrinsic periodicity of $\Delta N_a$ and also at twice that periodicity, with predictable ratios between their amplitudes and phases.
 
In practical terms, while the scaling dependence of $\alpha(\lambda)$ is computed directly from modeling ({\it cf.} Figure~\ref{fig:treerings}, upper left), it may also be estimated directly by comparing tree ring distortion features in the flat fields taken for illumination with different wavelengths. We expect $\alpha \approx 0$ for $\lambda<600$nm, with a gradual rise toward longer wavelengths. Indeed, this general trend is seen in the PRNU dependence on wavelength {\it between} the regimes dominated by unrelated phenomena ({\it e.g.}, dead layers and fringing), where we identify some of the PRNU values as proxies for the variance in $\Delta I$~\cite{Frank:privcomm}.

\subsection{Device front-side implant fixed pattern effects}

The two fixed pattern effects described in this section arise from lateral field terms due to structure in the bound charge distribution near the front side. The distributions in question are the $p+$ implants that form the channel stops and apparently the blooming stop. When depleted, they form a linear density $-\lambda$ of distributed negative charge, a small distance $a$ into the Si bulk from the clocks. Neglecting the periodic arrangement of equipotential boundary conditions formed by the parallel clocks for now, imagine for now that the front side surface is rather an equipotential surface that is perfectly planar. Together with the planar equipotential of the backside window, the symmetry of the boundary conditions lend to the {\it method of images} often used in electrostatics problems. The equipotential planes of the frontside and backside surfaces are established by using the following arrangement: The $p+$ implant linear density ($-\lambda$, at $z=+a$) has an image linear charge density ($+\lambda$, at $z=-a$). This pair of linear densities together form a 2-dimensional dipole, translationally invariant along one axis perpendicular to the dipole moment. The moment $\vec{\xi}_\perp=-2a\lambda\hat{k}$ contributes to $\delta\vec{E}(\vec{x})$ an amount $\delta\vec{E}^{\xi}_\perp(\vec{x})=(2[\vec{\xi}\cdot\hat{r}_\perp]\,\hat{r}_\perp-\vec{\xi})/(2\pi\epsilon_0\epsilon_{Si}\,r^2_\perp)$. In total, identical two-dimensional dipoles are placed at positions $z=0,\,\pm 2t_{Si},\,\pm 4t_{Si},\,\pm 6t_{Si}\, \cdots \,\pm 2(n-1)t_{Si},\,\pm 2nt_{Si}$ for some appropriately chosen $n$. This effectively establishes symmetry on either side of the equipotential planes at $z=0$ and $z=t_{Si}$. The field contribution from this dipole lattice arrangement is therefore:
\begin{eqnarray}
\delta\vec{E}_\perp(\vec{x}|\vec{x}_0)&=&{1 \over 2\pi\epsilon_0\epsilon_{Si}}\sum^{n}_{\ell =-n}{2[\vec{\xi}\cdot\hat{r}_{\ell\perp}]\hat{r}_{\ell \perp}-\vec{\xi}\over r^2_{\ell\perp}},\\
\vec{r}_{\ell \perp}&=&\vec{x}_\perp-(\vec{x}_{0\perp}+2\,\ell \,t_{Si}\,\hat{k}); \;\;\hat{r}_{\ell \perp}\equiv{\vec{r}_{\ell \perp}\over |\vec{r}_{\ell \perp}|}
\end{eqnarray}
where a vector's projection into two dimensions is given by $\vec{v}_\perp = \vec{v} - (\vec{v}\cdot\hat{\mathrm{e}})\,\hat{\mathrm{e}}$, where $\hat{\mathrm{e}}$ is the unit vector along the linear charge distribution axis of symmetry. In the model calculations that follow, field lines were traced using the perturbed electric field in order to locate both the saddle point loci and the points at the backside surface that map to those loci. The resulting distortion in the flat field response were compared to the data, and the self-consistent astrometric errors were also produced.

\subsubsection{Midline charge redistribution}\label{sec:f}
An abrupt drop in the photo response of the final row (together with an increased photo response in the rows preceding it) is seen in all amplifiers of a specific 16-segment sensor. For typical settings, the last row commonly contains less than 30\% of the flat field level. Because parallel transfer is performed in either direction away from the location of that final row (during integration), the two adjacent rows belonging to opposing segments are referred to as the "midline". Because the total charge collected and clocked out in a single column is equivalent to the product of the flat field level with the number of rows, this effect is also thought to not be a quantum efficiency variation, but a simple mapping function of the pixel boundaries (and a resultant variation in pixel size there). Early ideas of this mechanism were readily formulated because the {\it midline} is a {\it blooming stop} -- essentially a {\it p+} implant in an {\it n-channel} device, and should therefore resemble an isolated channel stop laid out perpendicularly to and intersecting with the channel stops that delimit all of the sensor's columns. Presumably the fact that this single implant is isolated - can make its effectively bound charge density under depletion ($N_a$) have a net repulsive effect on conversions (electrons) being drifted toward the grid of potential wells at the channel. Figure~\ref{fig:fixedpattern} (upper left) provides a comparison between measurement and calculation, for the midline redistribution effect. In addition to the flat field response, the calculation provides other ancillary pixel data (astrometric error, shape transfer). The midline charge redistribution effect is less sensitive to wavelength than tree rings are, so any flat field suppression term $\alpha(\lambda)$ must be measured or derived appropriately.

\subsubsection{Edge rolloff}\label{sec:g}
\begin{figure}[tbp] 
\centering
\begin{tabular}{rr}
\includegraphics[height=.4\textwidth,angle=-90]{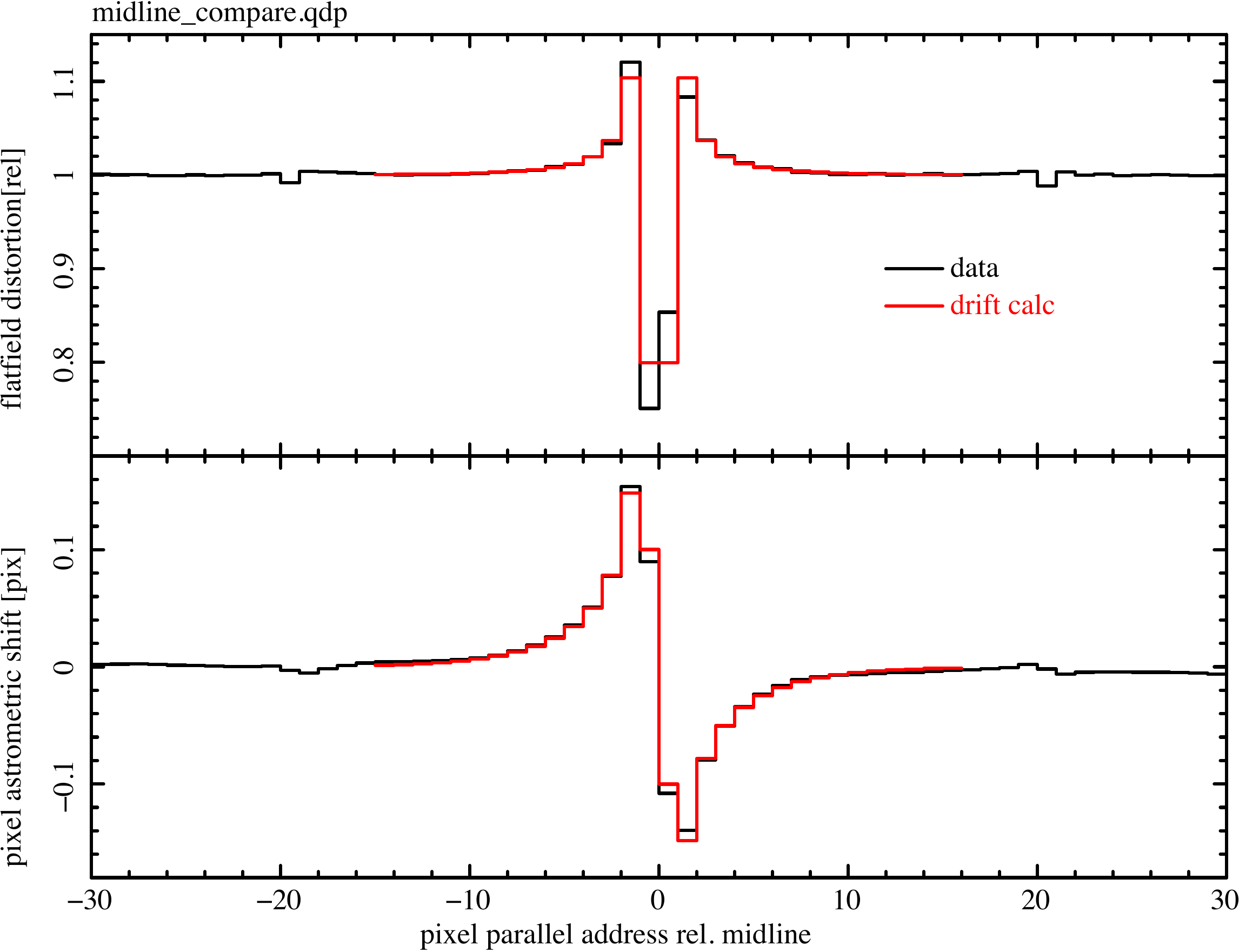}&
\includegraphics[height=.4\textwidth,angle=-90]{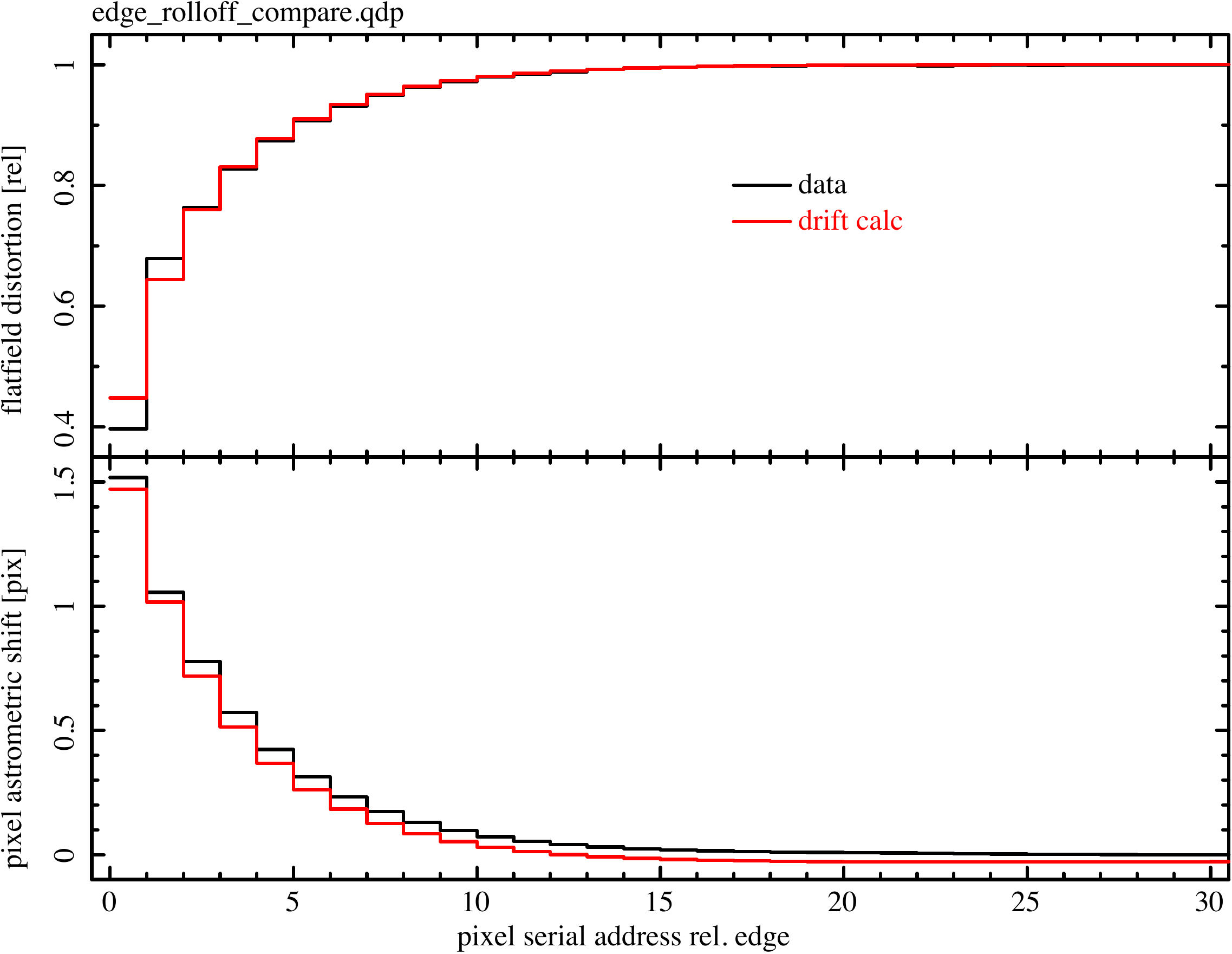}\\
\includegraphics[height=.4\textwidth,angle=-90]{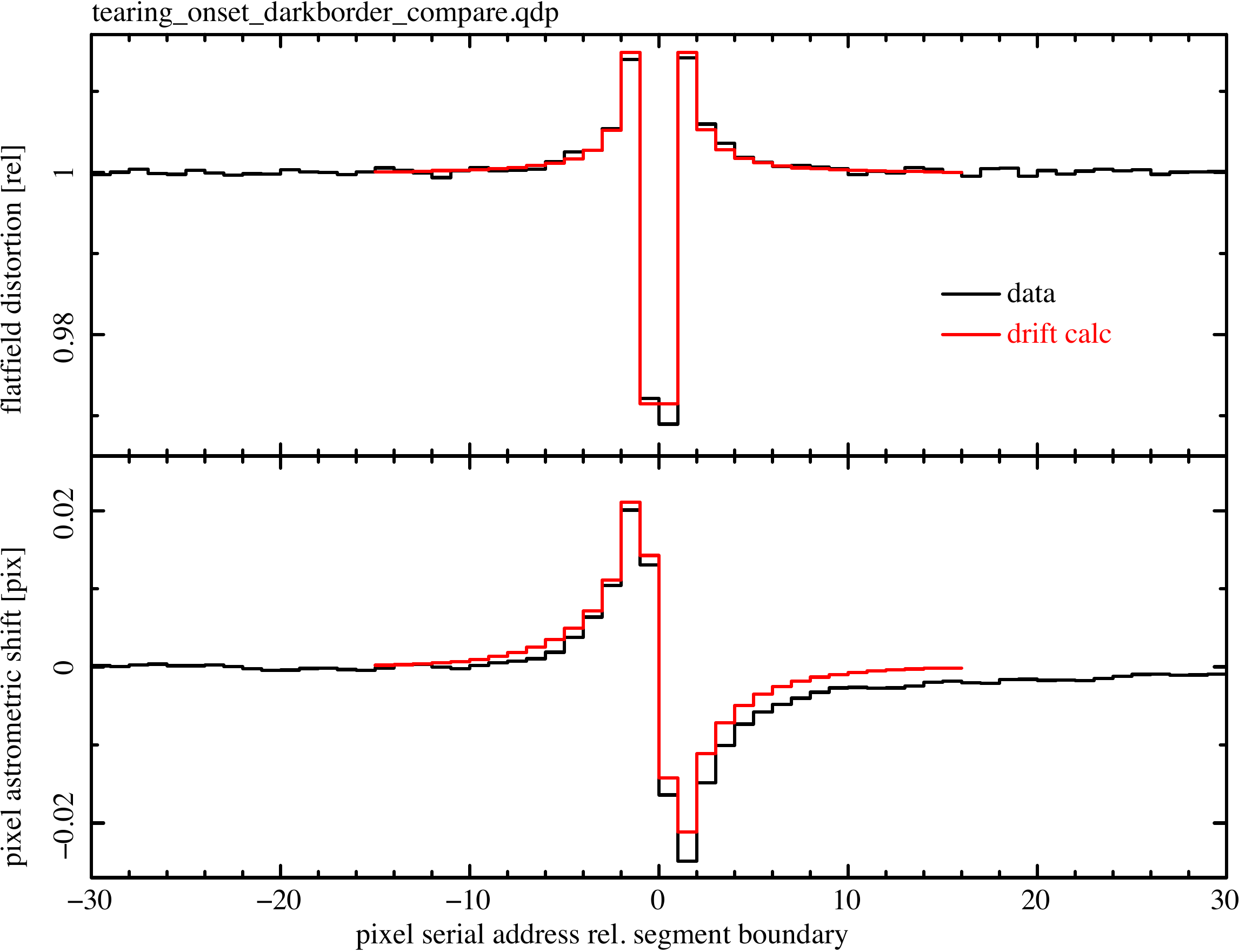}&
\includegraphics[height=.4\textwidth,angle=-90]{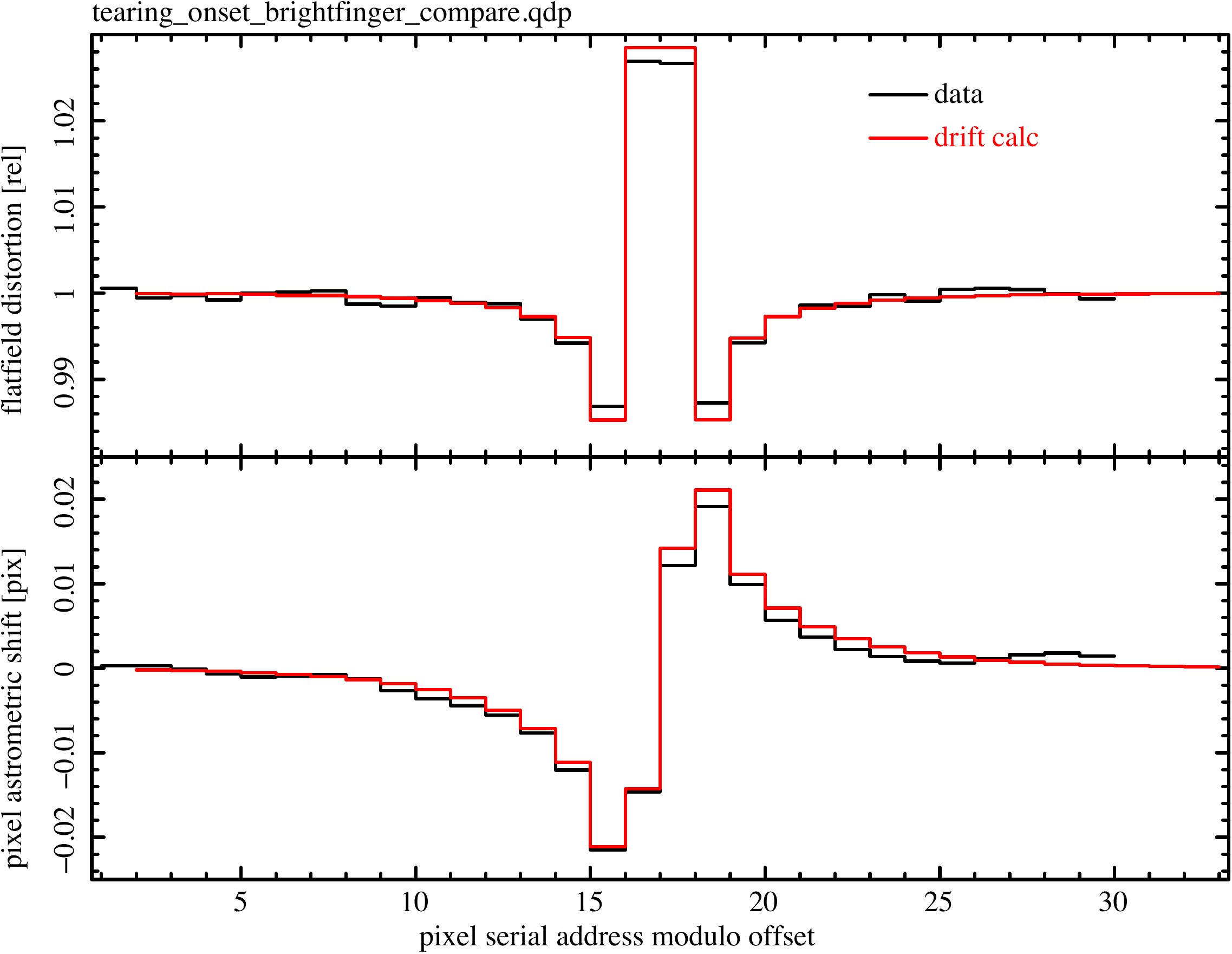}\\
\end{tabular}
\caption{Comparison of flat field response distortions in the data (black curves) to results of the drift calculation (red curves). While the drift calculation has access to the "truth" in reporting astrometric shifts, these shifts are estimated on the data side by assuming an astrometric boundary condition and that pixel response is affected by pixel size only. All dipole moment parameters for the $p+$ implants are quoted in units of $\xi_0 \equiv 2a\lambda_0 = 10^{-6}\,\mathrm{q}_{e}$. Upper left: the midline charge redistribution, modeled with $\vec{\xi}_\perp=-7\xi_0\,\hat{k}$. There appears to be a phasing error in the position of the bloom stop with respect to the integrating clocks. Upper right: the edge rolloff, modeled with a truncated array of channel stops, each with $\vec{\xi}_\perp=-28\xi_0\,\hat{k}$ and no guard ring. The difference between the data and model in the first two columns may in fact be most responsive to whether the guard ring drain is used or not. Lower left: "dark boundary" seen at the segment boundary, spanning amplifier segments. Lower right: a "bright finger" transient feature seen in the middle of a segment. Note the resemblance of these features with the midline feature in the upper left. These latter two features were modeled with $\vec{\xi}_\perp=-1\xi_0\,\hat{k}$ and $\vec{\xi}_\perp=+1\xi_0\,\hat{k}$, respectively. Additional ancillary pixel data resulting from these calculations include pixel shape transfer $\Delta S_{i,i+1}$ and diffusion scale $\sigma(z)$, which is in general position dependent. In the case of the edge rolloff, we find a local maximum in $\sigma(t_{Si})$ at the edge that is about 8\% larger than nominal in the edge-most pixels, corresponding to drift times that are $\sim$17\% greater than nominal.}
\label{fig:fixedpattern}
\end{figure}

Columns near the edge of the sensor, and rows nearest the serial registers, register a smaller signal than the flat field levels of the interior region. This trend is indeed monotonic and the rolloff is seen further into the imaging device than the sensor is thick (10 pixels). Proximity of these pixels to the guard ring suggest a dependence on three bias voltages: guard ring drain bias, integrating clock rail, and backside bias. For typical settings, edge column signal is near or below the 50\% level. Figure~\ref{fig:fixedpattern} (upper right) provides a comparison between measurement and calculation for the edge rolloff effect seen in the prototype sensors. In performing the calculations, the guard ring was neglected because its location and size are unknown.  Modest ($\sim$10\%) flat field distortion dependence on guard ring drain bias shall be used in the future to improve this model.

Recent spot projector measurements that show specific photometric results that differ from the flat field response~\cite{O'Connor:thisvolume} may be compared qualitatively to the predictions of Figure~\ref{fig:fixedpattern} (edge rolloff and midline charge redistribution). A detailed comparison would require a reasonably accurate model for the spatial distribution of the projected spots.

\subsubsection{Anti-correlation of adjacent pixels}\label{sec:h}
This topic is discussed in the study by Smith \& Rahmer~\cite{Smith:2008} and explored further by Kotov et al.~\cite{Kotov:2010}. They applied several methods to disentangle structure in the PRNU between QE and pixel area contributions, for two dissimilar sensors~\cite{Smith:2008} and for another thick sensor~\cite{Kotov:2010}. In all cases they found that a dominant contribution to the overall PRNU is attributed to variations in the boundary between {\it adjacent rows}, while variations in boundaries between {\it adjacent columns} can only be responsible for a much smaller contribution. While this apparent aperiodicity between some combination of the collection and barrier clocks (gates) falls squarely under the subject of our paper, there is probably not enough diagnostic information yet to accurately attribute this to geometric, lithographic process errors, a source of phasing errors in an otherwise perfectly periodic electrostatic (Dirichlet) boundary conditions at the front side oxide-silicon interface, some spatial non-uniformity of dopant at the channel, or some combination of the above. The one-dimensional ancillary pixel data analogues described in Eqs.~4.1-4.3 above may be applied to the pixel height jitter. We immediately see from these that for uncorrelated row boundary errors, the corresponding pixel astrometric jitter is simply proportional to the pixel flat field response jitter with $\sigma_P \approx {1\over 2} \sigma_I$ pixels. Similar considerations for tree ring induced errors predict that $\sigma_P \approx (p / 2 \pi) \sigma_I$ pixels (with period $p$ measured in pixels). We suggest that for astrometric concerns (e.g., image stacking, multi-fit, downstream shape estimation), the uncorrelated row boundary error induced PRNU effects amount to a smaller problem than the other, spatially correlated mechanisms will cause. 

For the purpose of this paper, we do not investigate this phenomenon further, except to note that the stark contrasts seen~\cite{Smith:2008}  between the conventional and the thick, fully depleted sensors' {\it smoothing rate index} -- essentially the reduction in computed variance as larger numbers of rows or columns are summed together  -- aren't adequately understood in our estimation. While these are not directly addressed here, we point to \S\ref{sec:k} as another mechanism that, in the case of thick sensors, may efficiently {\it complicate} the otherwise simple, shot noise-driven statistical predictions.

\subsection{Flux dependence of the fixed pattern features}\label{sec:i}
\begin{figure}[tbp] 
\centering
\begin{tabular}{rr}
\includegraphics[height=.4\textwidth,angle=-90]{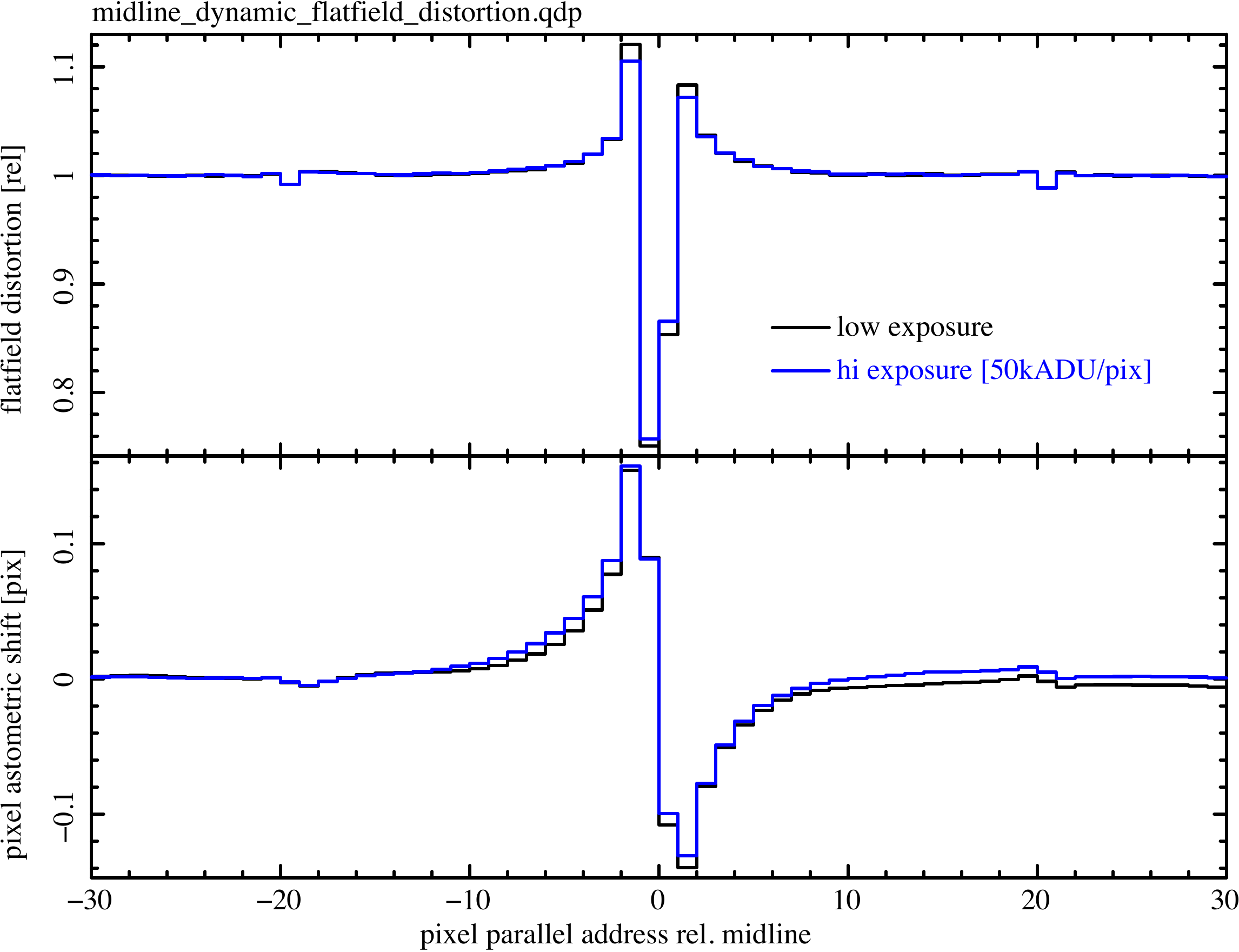}&
\includegraphics[height=.4\textwidth,angle=-90]{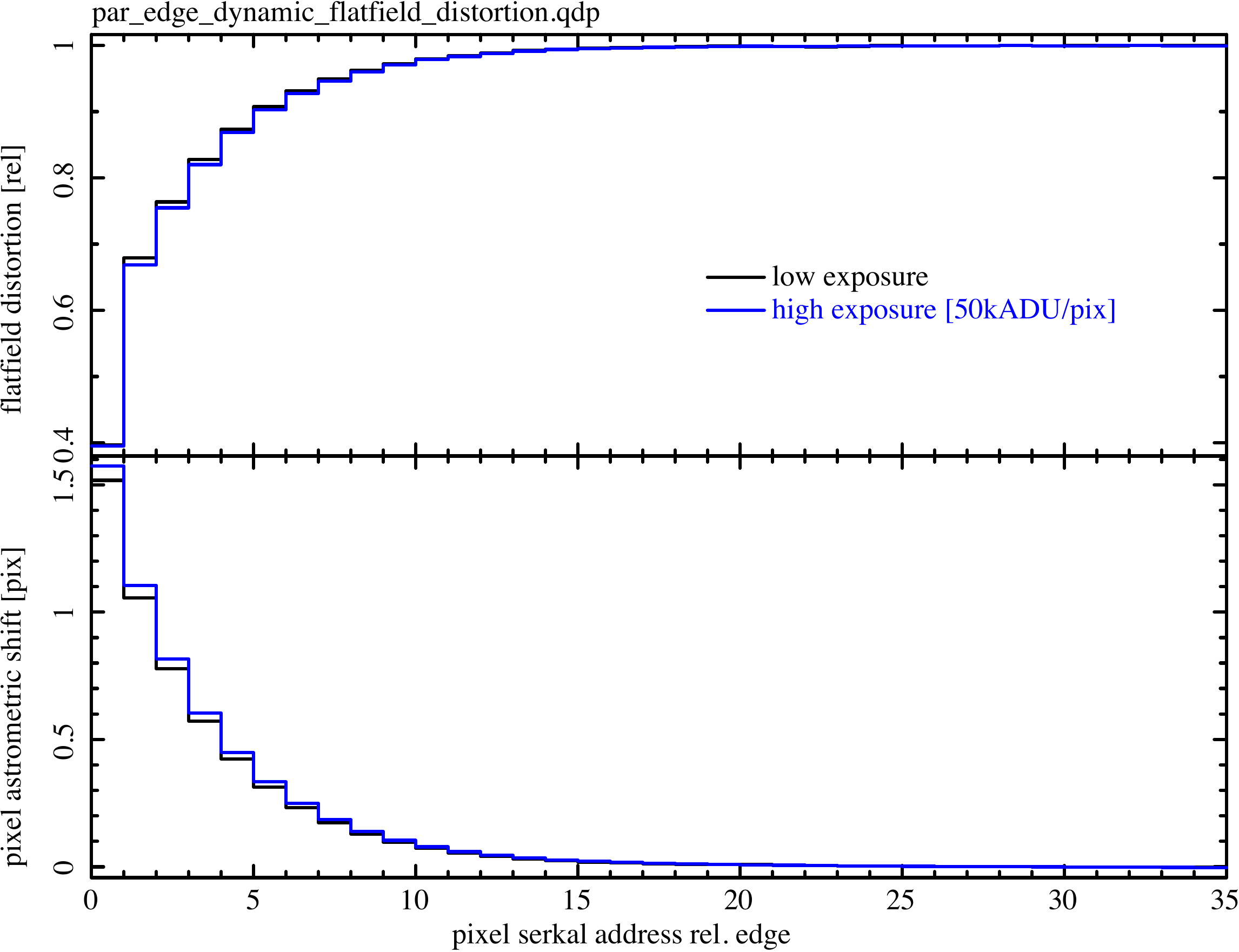}\\
\end{tabular}
\caption{We see definite variation in the flat field distortions due to the number of accumulated conversions collected at the channel. This is a coherent trend seen over a series of exposure levels, but only two cases are shown here for clarity. In the above plots, we compare the low-flux distortions (black curves) to the high-flux distortions due to approximately $10^{5}$ conversions per pixel (blue curves). Left: the midline redistribution; Right: the edge rolloff. In the midline case, adjacent pixels have a contrast of up to 3E4 electrons and we see a 2\% drop in pixel size for those pixels with the most response. In the rolloff case, presence of the charges collected at the channel results in a reduction in the pixel sizes near the edge, and an increased astrometric error for those pixels. These data sets provide unique opportunities to fine-tune the drift model to include charge clouds residing in the channel.}
\label{fig:dynamicfixedpattern}
\end{figure}

As mentioned briefly above, strict categorization of the distortion effects seen in the flat field response as either {\it fixed pattern} or {\it dynamic} is artificial. Further examination of the fixed pattern features described above reveal that these are also dynamic. Figure~\ref{fig:dynamicfixedpattern} shows dependence on the flat field level, or the number of conversions already collected at the channel when integration is halfway complete. In the context of the three distributions under consideration (illumination, static bound charge and collected conversions), the trend with increased exposure makes intuitive sense: A quantitative softening of the midline redistribution and a sharpening of the edge rolloff feature.

\subsection{"Tearing" onset - variation in hole occupancy within the channel stop array}\label{sec:j}
Coherent, transient features in the flat field response that resemble contours have been seen at the $\Delta\ln(C)\sim 0.05$ level. The "contours" form on the interior region of the sensor, notionally centered on the region with greatest illumination and extend across segments in a properly reassembled image that displays the sensor format. Details of the contour shape show "scalloped" features coincident with segment boundaries, but otherwise do not seem to be affected by the direction of serial transfer. The "contour" seen in the flat field response, on closer inspection, appears to have a bimodal response, with too much signal clocked in the parallel direction several rows before too little signal is clocked in the same direction. Sequences of images acquired under identical conditions typically show a significant change between the first and second image in the series, suggestive that sensor illumination history may be a "knob" on the effect. Careful inspection of many images exhibiting the contour (or "tearing") feature have revealed curious features in the flat field response, and provide what may be "calibration points" for what is at the heart of this phenomenon.  These include pairs of neighboring columns within a segment, that have excess photo-response ($\Delta\ln C \sim +0.05$) that extend from the first row, but ending abruptly at some higher parallel address. Similarly, pairs of adjacent columns straddling the segment boundary show a deficit in photo-response ($\Delta\ln C \sim -0.05$) - also extending from the first row to some higher parallel address value. In the latter case, the photo-response deficient columns appear to either increase in number (of columns) with parallel address, {\it or} the apparent response deficit would vanish at the row where the "tearing" contour occurs. In both these cases ({\it bright finger} and {\it dark border}), the axis of symmetry is not the column, but the dividing line {\it between} columns - apparently the channel stop. These observations led to speculation that this behavior in the imaging region may be a result of different carrier concentration (holes) within the channel stops: they are effectively the potential wells that can attract and harbor those carriers with a long recombination time. The channel stops that delimit sensor segments can be different from channel stops within the segments because of the topology of adjacent serial registers: there may be an {\it escape route} within the {\it p+} implant for the holes leave the imaging section of the device. Other channel stops within the segment, on the other hand, may simply dead end at the serial register. Under this scenario, holes would efficiently transfer from one row to the next toward the serial register (out of phase with the electrons) until their space density reaches saturation, and cannot be transferred by the parallel clock swings during readout. The net effect here would be that some regions in the imaging section have channel stops containing holes at some saturated level, while other regions have channel stops that are truly depleted, devoid of holes. These extrema may be what is behind the two apparent "states" routinely seen in the flat field response of images affected by "tearing".  Indeed, the differing electrostatic properties of those channel stop states, when isolated and juxtaposed, can bend drift field lines to distort flat field response {\it in adjacent pairs of columns}. A consequence of this is that a sensor may be operating in the {\it saturated channel stops} state, where the only "tearing" related artifact is a pair of dim columns (missing about 5\% of the expected conversions) between every pair of adjacent segments with neighboring serial registers. The artifact qualitatively resembles the {\it midline charge redistribution} with a smaller amplitude, but affects seven times as many pixels according to the current (8x2) segmentation. Figure~\ref{fig:fixedpattern} (lower left and lower right) display these features and initial efforts to model the underlying drift field distortion.

\begin{figure}[tbp] 
\centering
\begin{tabular}{rr}
\includegraphics[width=.3\textwidth,angle=0]{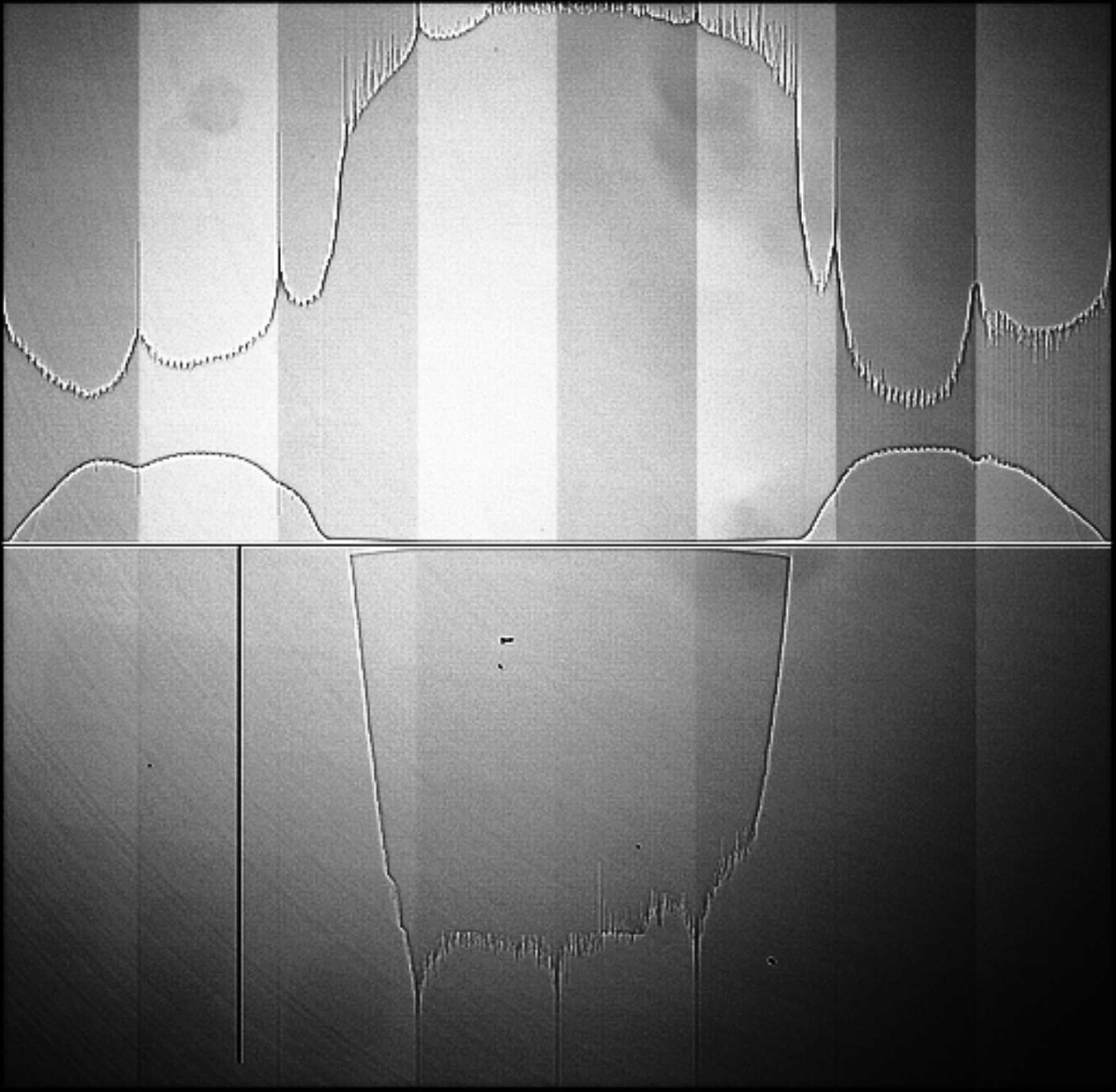}&
\includegraphics[width=.3\textwidth,angle=0]{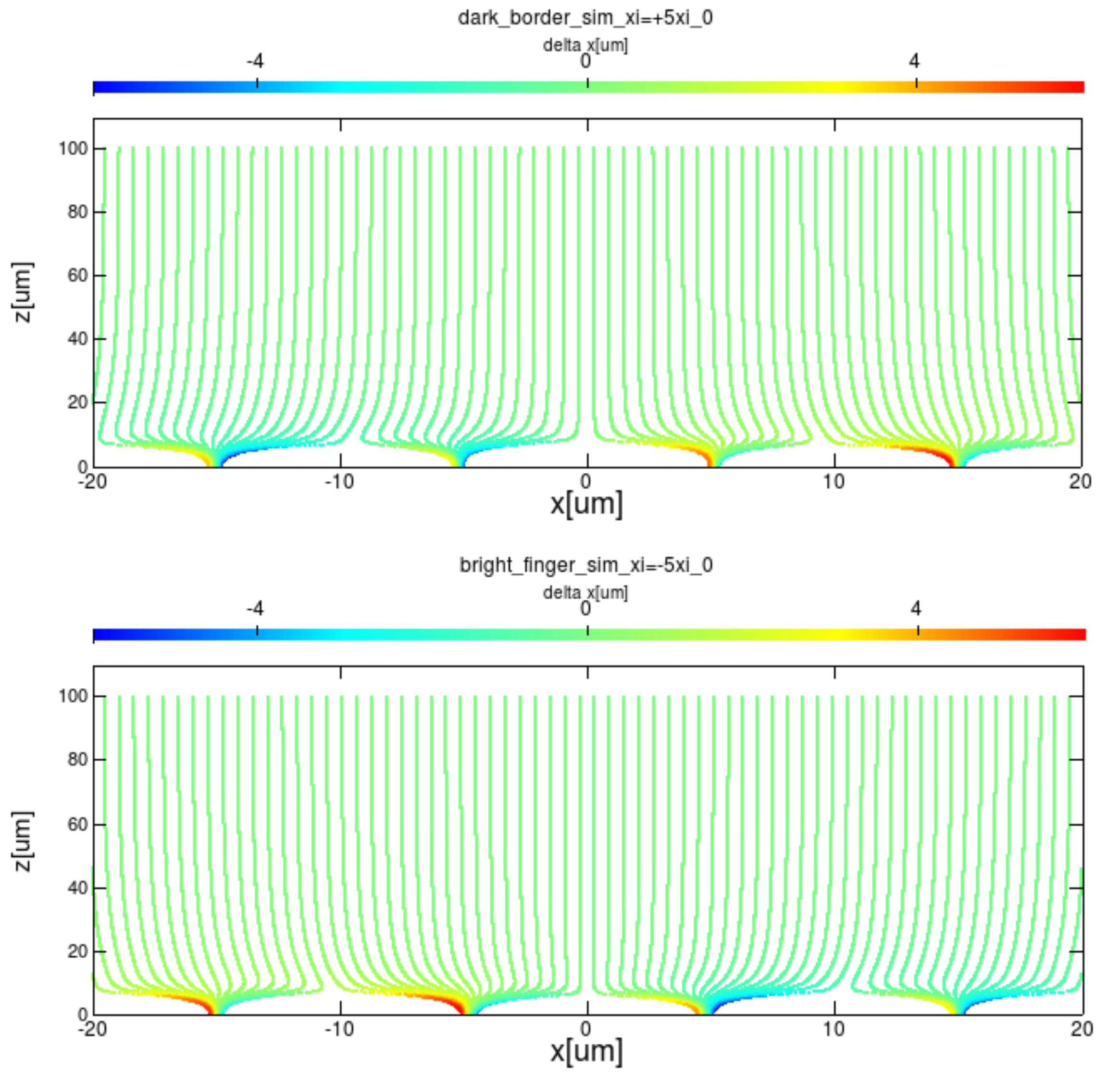}\\
\includegraphics[width=.3\textwidth,angle=0]{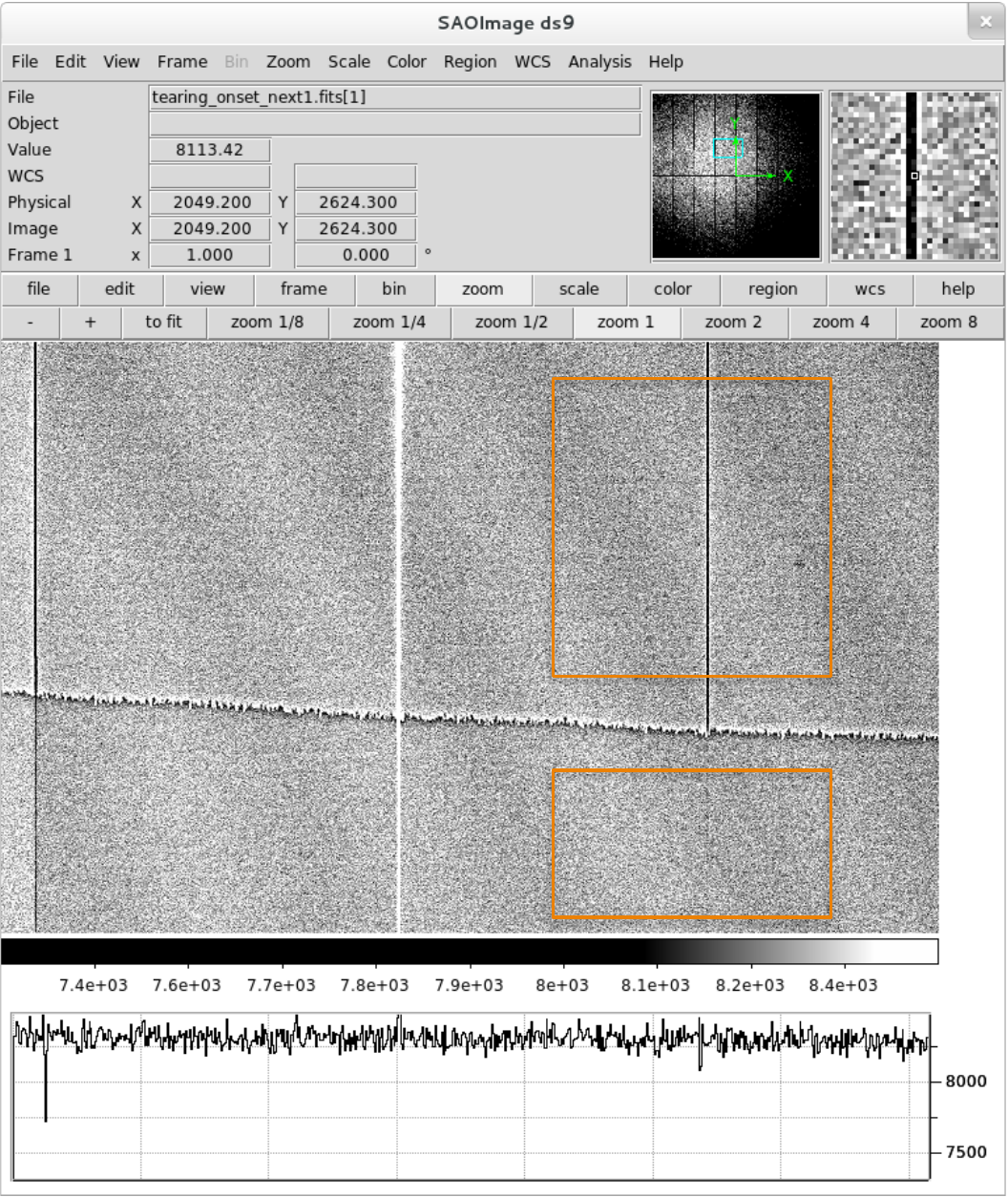}&
\includegraphics[width=.3\textwidth,angle=0]{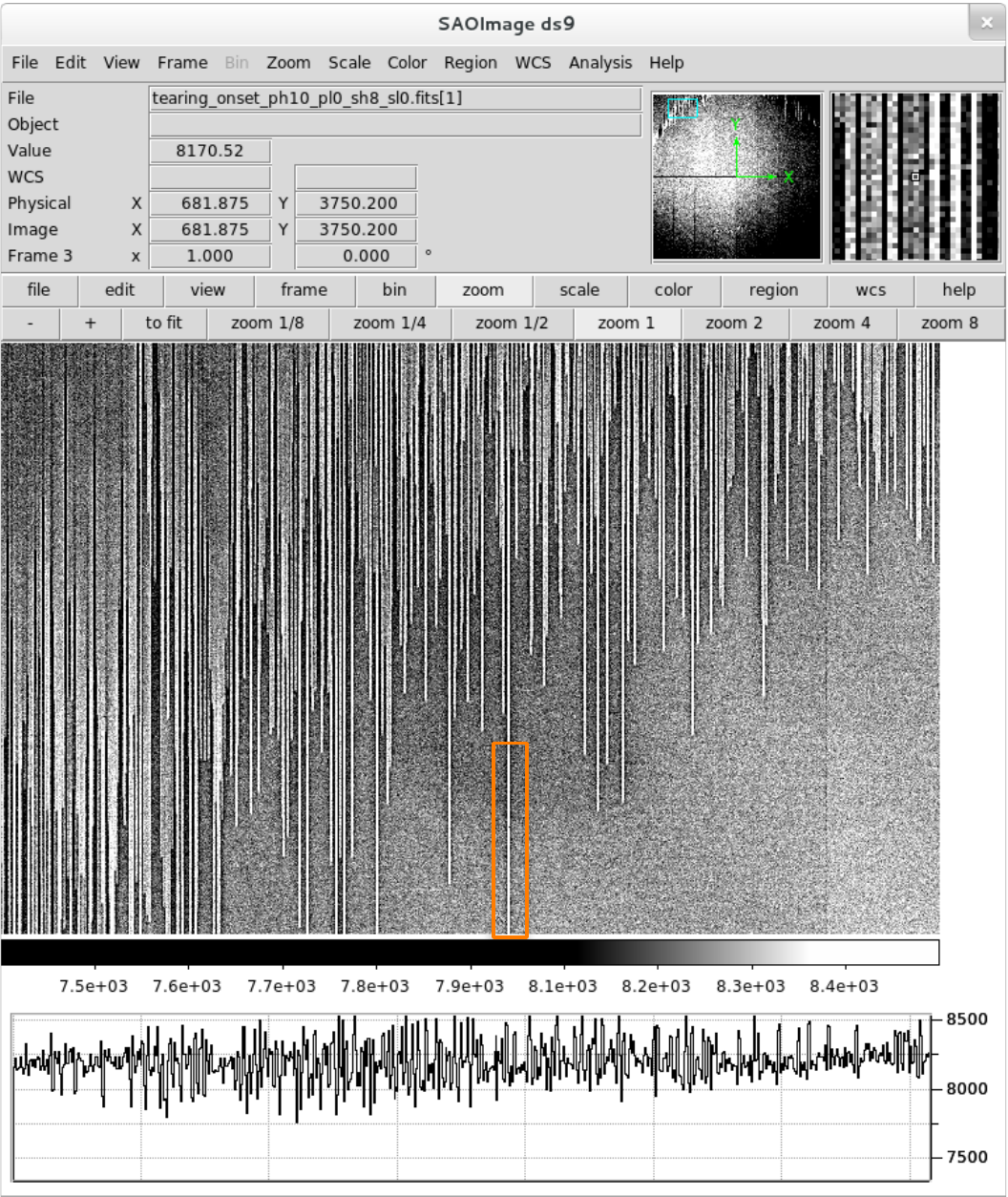}\\
\end{tabular}
\caption{Figures that show instances of the tearing features described in the text, together with exaggerated modeling calculations for clarity. Upper left: a sample full-format, reassembled image from a flat field exposure that shows a clear example of the bimodal contour described in the text. The stretch of this image is approximately $\Delta \ln C \approx 0.05$. The area circumscribed by the contour appears to increase abruptly after the first exposure of the sequence. Areas near the boundary that join the interior state with the exterior state exhibit strong flat field distortions with a net radial (outward) component to the drift field. Lower left: flat field response of the interior state is smooth, and shown in the lower rectangle, while the outer state shows bimodal interruptions (upper rectangle) surrounding the first and last columns of adjacent amplifiers (the "dark border"). Lower right: on rare occasions, images are acquired that appear to catch the tearing pattern while it is still equilibrating. This image shows an isolated pair of columns (the "bright finger") with an excess of collected conversions, surrounded by columns containing a deficit. For both panels on the bottom, the parallel transfer direction is toward the top of the page. Upper right: drift field calculations for the dark border (top) and the bright finger (bottom). These calculations are exaggerated by a factor of 5 as compared to the fits shown above so that the lateral drift effects can be discerned by eye. The single input parameter is the excess (dark border) or deficit (bright finger) in the equivalent channel stop 2-D dipole moment $\vec{\xi_\perp}$ located at $(x,y)=(0,0)$. In this case, the backdrop, periodic nominal channel stop dipole strength is $-40\xi_0\hat{k}$ with the central channel stop varying between $-45\xi_0\hat{k}$ (dark border) $-35\xi_0\hat{k}$ (bright finger). Although it is difficult to see in these images, the position of the saddle point between the central two pixels should be nearer the channel for the bright finger than for the dark border. The false color scale used in these plots indicate the integral of the lateral drift after launch at $z=100\mu$m.}
\label{fig:tearing}
\end{figure}

\subsection{Effects associated with pixel-to-pixel variation in conversions collected at the channel}\label{sec:k}

To approach the "brighter-fatter" effect\footnote{The observation that brighter sources tend to have systematically broader recorded widths than do fainter ones} using the tools described here, we include drift field distortions due to confined charge distributions to represent conversions collected at the channel. For the same reasons outlined above in the discussion of the linear charge density distributions, the {\it method of images} is used to describe the accumulated conversions as a dipole (in three dimensions this time), and then an expansion of that dipole's images are arranged on either side of the equipotential planes in order to satisfy the boundary conditions of this charge arrangement. Suppose the buried channel is a distance $z_{ch}$ from the front side surface, and the channel contains a charge value of $-N\mathrm{q}_e$. This charge packet has an image described by $+N\mathrm{q}_e$ located at $z=-z_{ch}$. The dipole moment $\vec{p}=-2z_{ch}N\mathrm{q}_e\,\hat{k}$ located at $z=0$. All dipole moments corresponding to channel occupancy are quoted in units of $p_0 \equiv 2z_{ch}N\mathrm{q}_e = 10^{5}\,\mu\mathrm{m}\,\mathrm{q}_e$. Adding up the dipole fields from the dipole image lattice is then:
\begin{eqnarray}
\delta\vec{E}(\vec{x}|\vec{x}_0) &=& {1 \over 4\pi\epsilon_0\epsilon_{Si}} \sum^n_{\ell=-n} { 3\left[ \vec{p}\cdot \hat{r}_\ell\right] \hat{r}_\ell - \vec{p} \over r_\ell^3}\\
\vec{r}_\ell&=&\vec{x}-(\vec{x}_0+2\,\ell\,t_{Si}\,\hat{k});\;\;\hat{r}_\ell\equiv{\vec{r}_\ell\over \left|\vec{r}_\ell\right|}.
\end{eqnarray}

\begin{figure}[tbp] 
\centering
\begin{tabular}{rr}
\includegraphics[height=.4\textwidth]{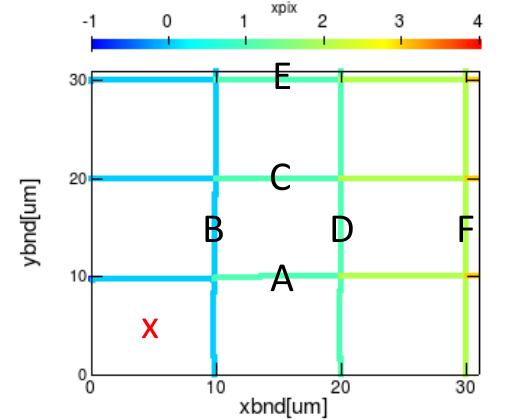}&
\includegraphics[height=.4\textwidth]{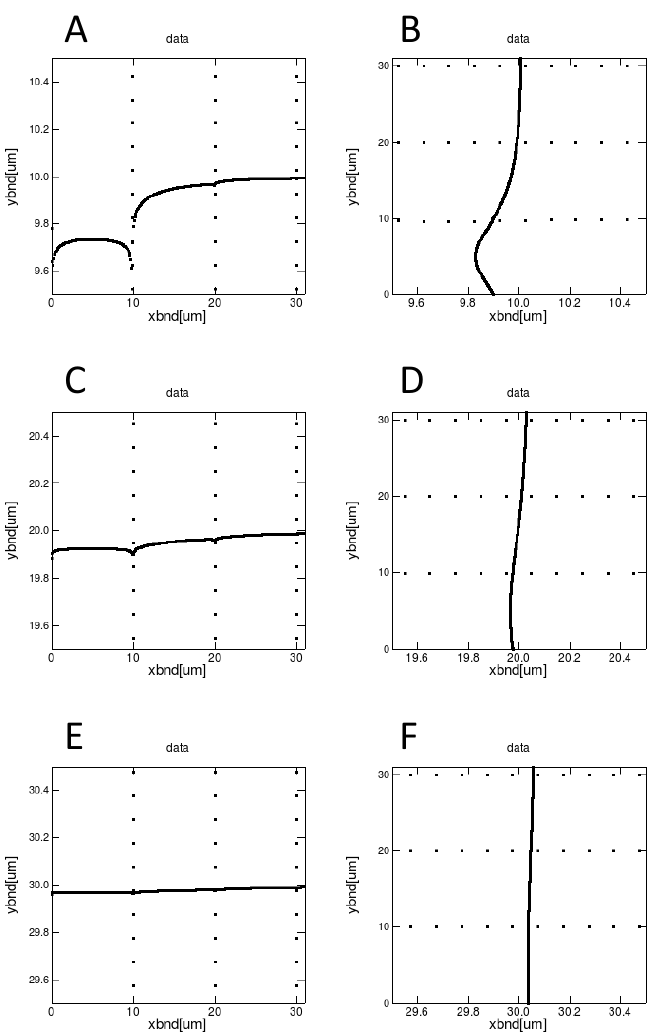}\\
\end{tabular}
\caption{Drift calculation results for pixel boundaries for the presence of an accumulation of conversions in a single pixel, labeled by the red "X" on the left hand side diagram. The distortion to the drift field was calculated for an assumed dipole strength of $\vec{p}=-p_0\,\hat{k}$. The right hand side displays six blow-ups of the pixel boundaries, labeled correspondingly A through F. The greatest distortions are of course seen at the boundaries that are only half a pixel from the occupied channel, those labeled A and B. The boundaries of the central pixel, mapped to the backside surface, are no longer square, nor rectangular -- a consequence of the {\it saddle point loci} corresponding approximately to the undistorted pixel boundaries. We divide the border distortion map by the input dipole moment to derive position dependent drift coefficients (in this case for surface conversions. The central pixel containing the red "X" has average drift coefficients of 0.015 and 0.030 pix$^2/p_0$ ($1\,\mathrm{pix}^2/p_0 \equiv 10^{-5} (z_{ch}/\mu\mathrm{m})^{-1}\,\mathrm{pix}^2/\mathrm{q}_e=10^{-7}\,\mathrm{cm}/\mathrm{q}_e$) along the serial and parallel directions, respectively. Note that finite drift coefficients exist for pixel boundaries that are up to several pixels distant from the aggressor.}
\label{fig:pixbndmap}
\end{figure}

\begin{figure}[tbp]
\begin{tabular}{rr}
\includegraphics[width=.4\textwidth]{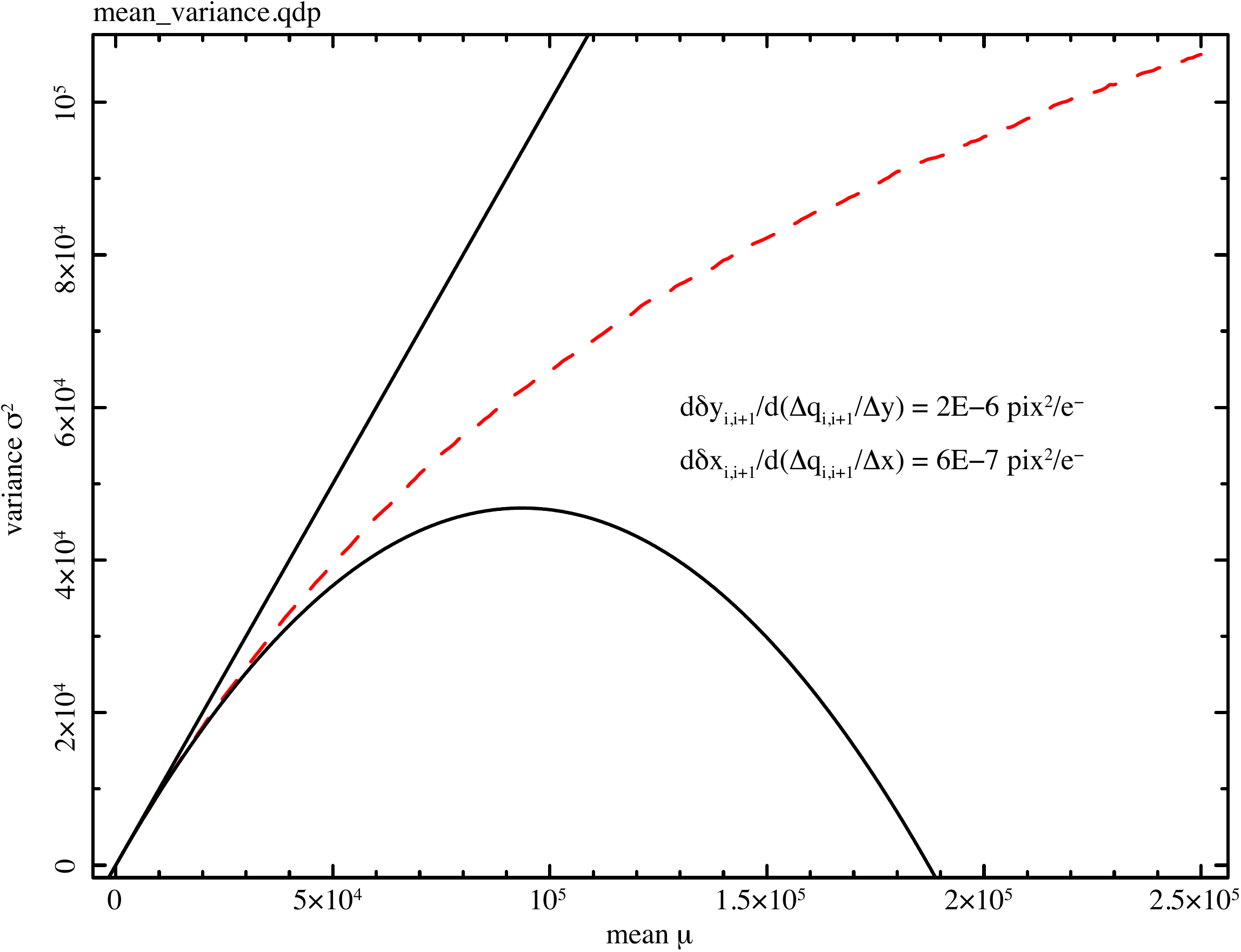}&
\includegraphics[width=.4\textwidth]{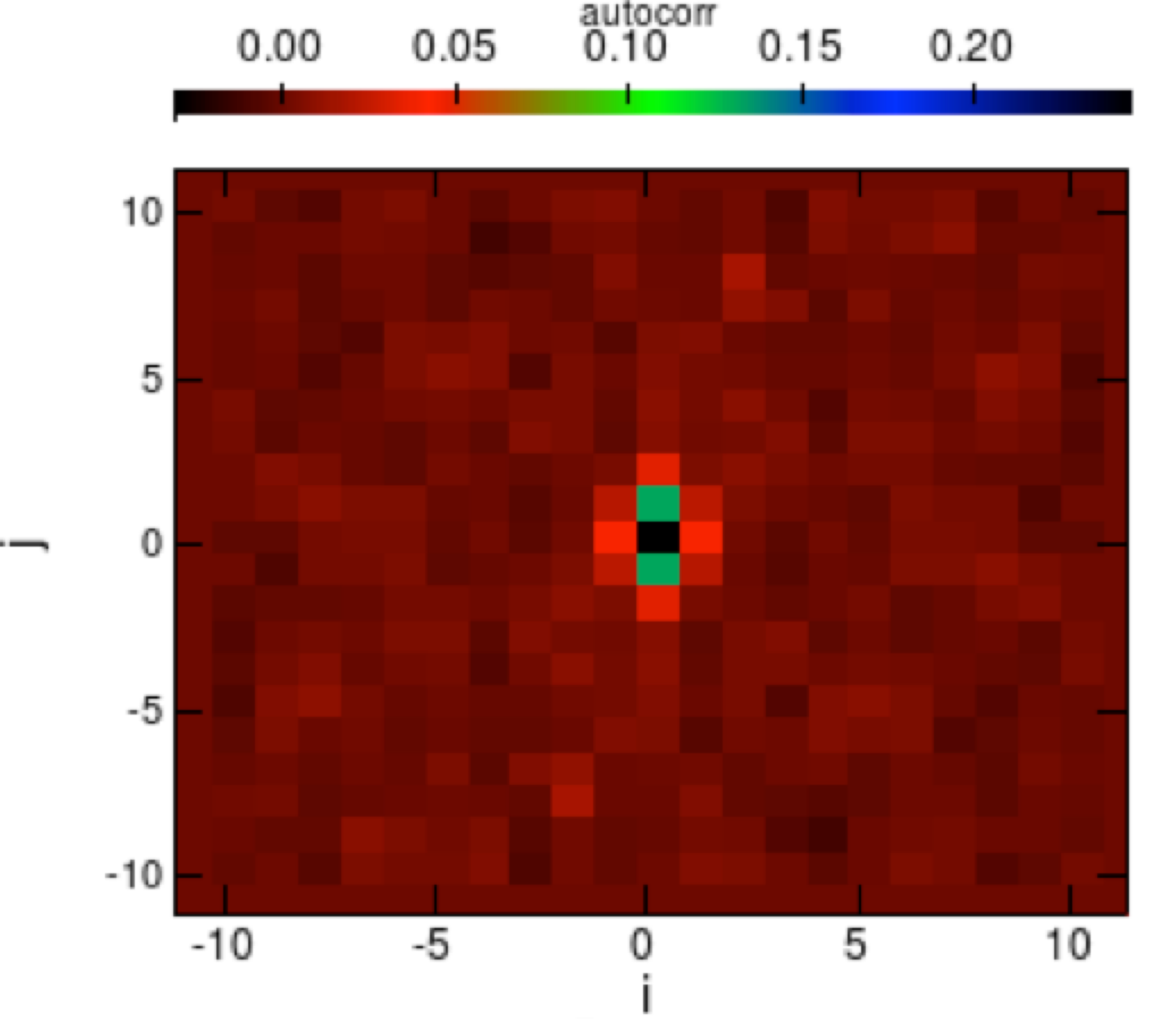}\\
\end{tabular}
\caption{Pixel response distortion predictions that are easily measurable with laboratory flat field data. Using two direction-specific drift coefficients shown, we performed a Monte-Carlo calculation of the {\it isotropic} mean-variance relation due to pixel boundary response to shot noise statistical fluctuations (red dashed curve on left). Solid black straight line there corresponds to the Poisson prediction while the parabolic, linear+quadratic fit made to the points with the smallest signal, provides a poor representation of the overall behavior. The quadratic coefficient happens to equal twice the sum of the two drift coefficients. Using the resulting two-dimensional response distribution evaluated for the maximum mean value ($2.5\times 10^5\,\mathrm{e}^-$), the autocorrelation is computed (right hand side) using the method described by P.~Astier~\cite{Astier:thisvolume}. This map displays the intrinsic anisotropy of the pixel boundary drift coefficients. It is currently unclear how this autocorrelation map can be used to predict autocorrelations at other flux levels due to significant departures from Poisson behavior.}
\label{fig:nonlinearities}
\end{figure}

Using this expression, we map the pixel boundaries back-projected onto the backside surface in the presence of some finite number of collected conversions at the channel, together with the perturbations imparted by the channel stop array and the periodic arrangement of integrating and barrier clocks (gates) along the orthogonal axis. The barrier clocks are seen to provide a significantly weaker confining field as compared to the channel stops. These results are shown in Figure~\ref{fig:pixbndmap} for surface conversions and clearly predict distortions to the pixel shape there, together with equivalent drift coefficients due to channel occupancy gradients as well as the response distortions shown in Figure~\ref{fig:nonlinearities}. We see the shape of a square pixel, distorted purely by its own collected charge, become convex along the parallel direction and concave along the serial direction with rotated corners. A complete description of this pixel boundary distortion response can be thought of as a Greens function, a fundamental property of the sensors. An arbitrary charge configuration can be used -- together with the Greens function -- to  efficiently update the pixel boundaries for new, incoming photo-distribution. 

\section{Summary and outlook}

Using a physical model for a high resistivity, thick, fully depleted CCD sensor, we computed what we believe are intrinsic drift field line distortions and corresponding drift coefficients for points within the Si bulk that map to the saddle point loci that form the depth specific pixel boundaries according to $\delta \vec{x}_\perp(\vec{x}_0|\vec{x}_{sp})$. The drift coefficients are scaled appropriately to match laboratory characterization data that exhibit flat field response distortions, and consistent ancillary pixel data (up to six values per pixel) are computed. Ancillary pixel data -- when computed and utilized properly -- would smooth out flat field exposures, correct astrometric errors, and cancel pixel elongation and shape transfer effects (each to first order), and thereby remove dominant instrument signature features intrinsic to the raw data. Modifications to catalog population algorithms, image stacking, and multi-fit strategies are foreseen. We see the use of ancillary pixel data $\Delta I_{i,i+1,j,j+1}$, $\Delta P_{i,i+1,j,j+1}$ and $\Delta S_{i,i+1,j,j+1}$ as a scientifically prudent and affordable option: an alternative to masking off, or specially treating, the data from significant fractions of the focal plane.\footnote{For the current segmentation geometry, focal plane losses are roughly 10\%, if affected regions are all 20 pixels wide} Once best guess first order errors have been accounted for, residual flat field response distortions may be correlated against residual astrometric error fields to assess second order corrections of these types.

\acknowledgments
Over the course of formulating this work, we enjoyed a wealth of insightful discussions with Pierre Antilogus, Pierre Astier, Steve Holland, Paul Jorden, Robert Lupton, Roger Smith and Chris Stubbs. High quality characterization data for LSST prototype sensors were acquired by Peter Doherty and others at Harvard, and by Paul O'Connor and his team at Brookhaven. LSST project activities are supported in part by a Cooperative Agreement with the National Science Foundation managed by the Association of Universities for Research in Astronomy (AURA), and the Department of Energy. Additional LSST funding comes from private donations, grants to universities, and in kind support from LSSTC Institutional Members.
\bibliographystyle{JHEP}
\bibliography{precision_astronomy_arasmus_2013_jinst_submit}

\providecommand{\href}[2]{#2}\begingroup\raggedright\begin{thebibliography}{10}

\bibitem{Groom:1999}
D.~E. {Groom}, S.~E. {Holland}, M.~E. {Levi}, N.~P. {Palaio}, S.~{Perlmutter},
  R.~J. {Stover}, and M.~{Wei}, {\it {Quantum efficiency of a back-illuminated
  CCD imager: an optical approach}},  in {\em Sensors, Cameras, and Systems for
  Scientific/Industrial Applications} (M.~M. {Blouke} and G.~M. {Williams},
  eds.), vol.~3649 of {\em Society of Photo-Optical Instrumentation Engineers
  (SPIE) Conference Series}, pp.~80--90, Apr., 1999.

\bibitem{Rajkanan:1979}
K.~{Rajkanan}, R.~{Singh}, and J.~{Shewchun}, {\it {Absorption coefficient of
  silicon for solar cell calculations}},  {\em Solid State Electronics} {\bf
  22} (Sept., 1979) 793--795.

\bibitem{Prigozhin:1998}
G.~Y. {Prigozhin}, A.~{Rasmussen}, M.~W. {Bautz}, and G.~R. {Ricker}, {\it
  {Model of the x-ray response of the ACIS CCD}},  in {\em X-Ray Optics,
  Instruments, and Missions} (R.~B. {Hoover} and A.~B. {Walker}, eds.),
  vol.~3444 of {\em Society of Photo-Optical Instrumentation Engineers (SPIE)
  Conference Series}, pp.~267--275, Nov., 1998.

\bibitem{Smith:2008}
R.~M. {Smith} and G.~{Rahmer}, {\it {Pixel area variation in CCDs and
  implications for precision photometry}},  in {\em High Energy, Optical, and
  Infrared Detectors for Astronomy III}, vol.~7021 of {\em Society of
  Photo-Optical Instrumentation Engineers (SPIE) Conference Series}, Aug.,
  2008.

\bibitem{Stubbs:thisvolume}
C.~{Stubbs}, ``Introduction and overview.'' Talk delivered at Precision
  Astronomy with Fully Depleted CCDs (contribution in this volume), Nov., 2013.

\bibitem{Canali:1975PhRvB}
C.~{Canali}, C.~{Jacoboni}, F.~{Nava}, G.~{Ottaviani}, and
  A.~{Alberigi-Quaranta}, {\it {Electron drift velocity in silicon}},  {\em
  Physical Review B} {\bf 12} (Sept., 1975) 2265--2284.

\bibitem{Canali:1975ApPhL}
C.~{Canali}, C.~{Jacoboni}, G.~{Ottaviani}, and A.~{Alberigi-Quaranta}, {\it
  {High-field diffusion of electrons in silicon}},  {\em Applied Physics
  Letters} {\bf 27} (Sept., 1975) 278.

\bibitem{Jacoboni:1977}
C.~{Jacoboni}, C.~{Canali}, G.~{Ottaviani}, and A.~{Alberigi Quaranta}, {\it {A
  review of some charge transport properties of silicon}},  {\em Solid State
  Electronics} {\bf 20} (Feb., 1977) 77--89.

\bibitem{Frank:privcomm}
J.~{Frank}, P.~{O'Connor}, and I.~{Kotov}, 2011.
\newblock Private communication, generic PRNU($\lambda$) measurements acquired
  on LSST prototype sensors.

\bibitem{O'Connor:thisvolume}
P.~{O'Connor}, ``Spot scan probe of edge and midline effects in fully-depleted
  ccds.'' Talk delivered at Precision Astronomy with Fully Depleted CCDs
  (contribution in this volume), Nov., 2013.

\bibitem{Kotov:2010}
I.~V. {Kotov}, A.~I. {Kotov}, J.~{Frank}, P.~{Kubanek}, M.~{Prouza},
  P.~{O'Connor}, V.~{Radeka}, and P.~{Takacs}, {\it {Study of pixel area
  variations in fully depleted thick CCD}},  in {\em High Energy, Optical, and
  Infrared Detectors for Astronomy IV}, vol.~7742 of {\em Society of
  Photo-Optical Instrumentation Engineers (SPIE) Conference Series}, July,
  2010.

\bibitem{Astier:thisvolume}
P.~{Astier}, ``The brighter-fatter effect and pixel correlations.'' Talk
  delivered at Precision Astronomy with Fully Depleted CCDs (contribution in
  this volume), Nov., 2013.

\end{thebibliography}\endgroup

\end{document}